\documentclass[a4paper,11pt]{article}
\pdfoutput=1 

\usepackage[utf8]{inputenc}

\usepackage[no-natbib-sort]{jheppub}
\usepackage{enumitem}

\usepackage{youngtab}
\usepackage{hyperref}
\usepackage{fancybox}
\usepackage{float}
\usepackage{amsfonts}
\usepackage{amsmath}
\usepackage{amsbsy}
\usepackage{graphicx}
\usepackage{amssymb}
\usepackage{amsthm}
\usepackage{bm}
\usepackage{comment}
\usepackage{epsfig}
\usepackage{latexsym}
\usepackage{pdflscape}
\usepackage{cancel} 
\usepackage{color}
\usepackage{cleveref}
\usepackage{bm,nicefrac}

\usepackage{xcolor}

\makeatletter
\def\@fpheader{\relax}
\makeatother

\DeclareSymbolFont{AMSa}{U}{msa}{m}{n}
\DeclareSymbolFont{AMSb}{U}{msb}{m}{n}
\DeclareMathSymbol{\fieldR}{\mathalpha}{AMSb}{"52}

\newcommand{\beq}{\begin{eqnarray}}
\newcommand{\eeq}{\end{eqnarray}}
\newcommand{\bea}{\begin{eqnarray}}
\newcommand{\eea}{\end{eqnarray}}
\newcommand{\be}{\begin{equation}}
\newcommand{\ee}{\end{equation}}
\newcommand{\bq}{\begin{equation}}
\newcommand{\eq}{\end{equation}}

\def\6{\partial}

\def\6{\partial}

\usepackage{color}
    \definecolor{darkgreen}{rgb}{0,0.5,0}
    \definecolor{darkblue}{rgb}{0,0,0.6}
    \definecolor{purple}{rgb}{0.4,.2,0.7}

\newcommand{\fig}[1]{Fig.~\ref{#1}}
\newcommand{\figs}[1]{Figs.~\ref{#1}}
\newcommand{\sect}[1]{Sec.~\ref{#1}}
\newcommand{\taut}{T}

\newcommand{\nc}{N}

\newcommand{\Sec}[1]{Sec.~\ref{#1}}

\def\E{{\mathcal{E}}}

\newcommand{\eqn}[1]{(\ref{#1})}
\newcommand{\eqq}[1]{Eq.~(\ref{#1})}

\newcommand{\Ts}{T_s}
\newcommand{\ts}{t_s}
\newcommand{\Es}{\mathcal{E}_s}
\newcommand{\Eh}{\mathcal{E}_\textrm{high}}
\newcommand{\El}{\mathcal{E}_\textrm{low}}
\newcommand{\Hs}{H}

\newcommand{\Tn}{T_n}
\newcommand{\tn}{t_n}

\newcommand{\Tg}{T_g}
\newcommand{\tg}{t_g}

\newcommand{\Trad}{T_\textrm{rad}}
\newcommand{\rhogw}{\rho_\textrm{GW}}

\newcommand{\kmax}{k_\textrm{max}}
\newcommand{\gmax}{\gamma_\textrm{max}}
\newcommand{\gtot}{\gamma_\textrm{tot}}

\newcommand{\kmaxhyd}{k_\textrm{max}^\textrm{hyd}}
\newcommand{\gmaxhyd}{\gamma_\textrm{max}^\textrm{hyd}}
\newcommand{\kstarhyd}{k_*^\textrm{hyd}}
\newcommand{\klow}{k_\textrm{low}}

\newcommand{\Ehigh}{\mathcal{E}_\textrm{high}}
\newcommand{\Elow}{\mathcal{E}_\textrm{low}}

\def\x{\pmb{x}}
\def\k{\pmb{k}}
\def\q{\pmb{q}}
\def\v{\pmb{v}}

\def\hq{{\tilde q}}
\def\hk{{\tilde k}}

\newcommand{\we}{\omega}

\newcommand{\Elat}{\mathcal{E}_\textrm{latent}}

\newcommand{\cc}{\cos \theta}

\newcommand{\sac}{\, , \qquad}

\title{Spinodal Gravitational Waves}

\author[a,b]{Yago~Bea,}
\author[b]{Jorge~Casalderrey-Solana,}
\author[c]{Thanasis~Giannakopoulos,}
\author[b]{Aron~Jansen,}
\author[d]{Sven~Krippendorf,}
\author[b,e]{David~Mateos,}
\author[b]{Mikel~Sanchez-Garitaonandia}
\author[c,f]{and Miguel~Zilh\~ao}

\affiliation[a]{Department of Physics and Helsinki Institute of Physics, PL 64, FI-00014 University of Helsinki, Finland.}
\affiliation[b]{Departament de F\'\i sica Qu\`antica i Astrof\'\i sica and Institut de Ci\`encies del Cosmos (ICC),  Universitat de Barcelona, Mart\'\i\  i Franqu\`es 1, ES-08028, Barcelona, Spain.}
\affiliation[c]{Centro de Astrof\'{\i}sica e Gravita\c c\~ao -- CENTRA,
  Departamento de F\'{\i}sica, Instituto Superior T\'ecnico -- IST, Universidade
  de Lisboa -- UL, Av.\ Rovisco Pais 1, 1049-001 Lisboa, Portugal }
\affiliation[d]{Arnold Sommerfeld Center for Theoretical Physics, Ludwig-Maximilians-Universit\"at,  Theresienstrasse 37, 80333 M\"unchen, Germany}
\affiliation[e]{Instituci\'o Catalana de Recerca i Estudis Avan\c cats (ICREA), Passeig Llu\'\i s Companys 23,  ES-08010, Barcelona, Spain.}
\affiliation[f]{Departamento de Matem\'atica da Universidade de Aveiro and
Centre for Research and Development in Mathematics and Applications (CIDMA), 
Campus de Santiago,
3810-183 Aveiro, Portugal}

\emailAdd{yagobea@icc.ub.edu}
\emailAdd{jorge.casalderrey@ub.edu}
\emailAdd{athanasios.giannakopoulos@tecnico.ulisboa.pt}
\emailAdd{a.p.jansen@icc.ub.edu}
\emailAdd{slk38@cam.ac.uk}
\emailAdd{dmateos@fqa.ub.edu}
\emailAdd{mikel.sanchez@polytechnique.edu}
\emailAdd{mzilhao@ua.pt}

\preprint{LMU-ASC 59/21}

\abstract{We uncover a new gravitational-wave production mechanism in cosmological, first-order, thermal  phase transitions. These are usually assumed to proceed via the nucleation of bubbles of the stable phase inside the metastable phase. However, if the nucleation rate is sufficiently suppressed, then the Universe may supercool all the way down the metastable branch and enter the spinodal region. In this case the transition  proceeds via the exponential growth of unstable modes and the subsequent formation, merging and relaxation  of phase domains. We use holography to follow the real-time evolution of this process in a strongly coupled, four-dimensional gauge theory and compute the resulting gravitational-wave spectrum.
We  discuss  the possibility that the spinodal dynamics may be preceded by a period of thermal inflation.}

\begin{document} 
\maketitle
\flushbottom

\section{Introduction}
\label{intro}
A first-order  phase transition in the Early Universe would produce Gravitational Waves (GW) that could be detected in current or future experiments. Both the deconfinement transition in Quantum Chromodynamics \cite{Aoki:2006we} at a scale $\sim 10^2$ MeV and the transition in the Electroweak sector \cite{Kajantie:1996mn,Laine:1998vn,Rummukainen:1998as} at $\sim 10^2$ GeV are actually smooth crossovers. Therefore, the discovery of GWs originating from a cosmological phase transition would amount to the discovery of new physics beyond the Standard Model (SM) of particle physics. Furthermore, in some cases this may be our only window into such physics. 

The sector responsible for the phase transition may range from an extension of the SM to a hidden sector coupled only gravitationally to the SM. In the first case, the  transition could take place at any scale between the Electroweak scale and the Plack scale $\sim 10^{19}$ GeV. For example, the Electroweak crossover  turns into a first-order phase transition even in minimal extensions of the SM  \cite{Carena:1996wj,Delepine:1996vn,Laine:1998qk,Huber:2000mg,Grojean:2004xa,Huber:2006ma,Profumo:2007wc,Barger:2007im,Laine:2012jy,Dorsch:2013wja,Damgaard:2015con},  resulting in a GW frequency in the mHz range  potentially observable by LISA \cite{Caprini:2019egz}. In the second case, the transition could take place virtually at any scale. For example, string theory compactifications often feature a large number of hidden sectors that are only gravitationally coupled to the SM degrees of freedom. Phase transitions in these sectors can lead to a sizable GW background~\cite{Schwaller:2015tja,GarciaGarcia:2016xgv,Huang:2020crf,Halverson:2020xpg} whose frequency can span the whole range of parameter space that will be explored by current and future GW detectors. This includes the high-frequency range $\gtrsim$ 30 kHz, where  new technologies are necessary \cite{Aggarwal:2020olq}, but also where conventional astrophysical foregrounds are absent.

Maximising the discovery potential requires an accurate understanding of the phase transition dynamics. This is usually assumed to proceed via the nucleation of bubbles of the stable, low-energy phase inside the metastable, supercooled phase~\cite{Hindmarsh:2020hop}. 
However, we will see in \Sec{thermo} that, if the nucleation rate is sufficiently suppressed, then the Universe will supercool all the way down to the end of the metastable branch
(see \cite{Albrecht:1982wi} for an early discussion of this possibility). This may happen if the number of degrees of freedom involved in the transition is large, as in gauge theories with a large number of colors. In this case the Universe will eventually enter the spinodal branch (see \fig{free}) and the transition will proceed via the exponential growth of unstable modes and the subsequent formation, merging and relaxation of phase domains. 
The purpose of this paper is to provide the first calculation of the resulting GW spectrum. In \Sec{thermo} we will see that, depending on the parameters of the model, the spinodal dynamics may or may not be preceded by a period of thermal inflation \cite{Lyth:1995hj,Lyth:1995ka}. 

To compute the GW spectrum, we must be able to follow the real-time evolution of the stress tensor of the system undergoing the phase transition. We will assume that this system is a strongly coupled, large-$N$, four-dimensional gauge theory. We will then use holography to map the quantum dynamics of the gauge theory stress tensor to classical gravitational dynamics in five dimensions. Our holographic model, together with the thermodynamics of the dual gauge theory,  will be described in \Sec{model}. In \Sec{time} we will determine the time evolution of the gauge theory stress tensor for an initial state on the spinodal branch.  In \Sec{gw} we will use the results from \Sec{time} to compute the GW spectrum. 

We emphasize that we are not simulating dynamical gravity in the four-dimensional spacetime where the gauge theory lives. We are simply using five-dimensional gravity to compute the time-evolution of the stress tensor of this gauge theory in flat Minkowski spacetime. In other words, holography is just a convenient tool to solve the quantum dynamics of the gauge theory. Once the stress tensor is known, one may forget about holography altogether and simply use the resulting stress tensor as a source for GWs in four dimensions. We will come back to this point in \Sec{disc}.
 
Holography has been previously used to study the real-time dynamics of the spinodal instability in a four-dimensional gauge theory in \cite{Attems:2017ezz,Attems:2019yqn} (see \cite{Janik:2017ykj} for a three-dimensional example). Because of technical limitations, these references imposed translational invariance along all but one of the gauge theory directions,  in such a way that the dynamics was effectively 1+1 dimensional in the gauge theory and 2+1 dimensional on the gravity side. Under these conditions, the transverse-traceless part of the gauge theory stress tensor vanishes identically and there is no production of GWs. Generating GWs requires at least 2+1 dimensional dynamics in a four-dimensional  theory, which  translates into  3+1 dimensional dynamics in the five-dimensional bulk. As is well known, numerically evolving the 3+1 dimensional Einstein equations is challenging and it could not be done with the code used in  our previous work  \cite{Attems:2017ezz,Attems:2019yqn,Attems:2016tby,Attems:2017zam,Attems:2018gou,Bea:2020ees,Bea:2021zsu,Bea:2021ieq}.  Fortunately, we have now developed a new code capable of performing 3+1 dimensional evolution \cite{code}, 
from which we will extract the corresponding spectrum of GWs.

Related 
work includes holographic applications to GWs \cite{Bigazzi:2020avc,Ares:2020lbt,Ares:2021nap} and to bubble dynamics \cite{Bigazzi:2020phm,Bea:2021zsu,Bigazzi:2021ucw,Ares:2021ntv}.

\section{The theory}
\label{model}
\subsection{Gravity model}
Our gravity model is described by the Einstein-scalar action
\begin{equation}
\label{action}
S=\frac{2}{\kappa_5^2} \int d^5 x \sqrt{-g} \left[ \frac{1}{4} {\cal R}  - \frac{1}{2} \left( \nabla \phi \right) ^2 - V(\phi) \right ]  \,,
\end{equation}
where $\kappa_5$ is the five-dimensional gravitational constant. Exclusively for simplicity, we assume that the scalar potential $V(\phi)$ can be derived from a superpotential $W(\phi)$ through the usual relation
\begin{equation}
 V(\phi)=-\frac{4}{3}W(\phi)^2+\frac{1}{2}W'(\phi)^2 \,.
\label{potentialsuperpotential}
\end{equation}
Different choices of (super)potential correspond to different dual four-dimensional gauge theories. As in \cite{Bea:2018whf,Bea:2020ees,Bea:2021zsu} we choose
\begin{equation}
\ell W(\phi)=
-\frac{3}{2}-\frac{\phi^2}{2}-\frac{\phi^4}{4 \phi_M^2}+\frac{\phi^6}{\phi_Q} \,,
\label{superpotential}
\end{equation}
where $\ell$ is the asymptotic curvature radius of the corresponding AdS geometry  and $\phi_M$ and $\phi_Q$ are constants, which in this paper we will set to 
\be
\label{param}
\phi_M=1 \sac \phi_Q=10 \,.
\ee
 The dual gauge theory is  a non-conformal theory obtained by deforming a conformal field theory (CFT)  with a dimension-three scalar operator $\mathcal{O}$ with source $\Lambda$. The latter is an intrinsic energy scale in the gauge theory that will set the characteristic scale of much of the physics of interest.  On the gravity side, $\Lambda$   appears as a boundary condition for the scalar $\phi$.  In the limit $\phi_Q \to \infty$ the sextic term in $W$ is absent and the model reduces to that in \cite{Attems:2017ezz,Attems:2019yqn,Attems:2018gou}.
 
The motivation for our choice of model is that it is possibly the simplest one with a number of desirable features. The presence of the scalar $\phi$ breaks conformal invariance. The first two terms in the superpotential are fixed by the asymptotic AdS radius and by the dimension of the dual scalar operator, $\mathcal{O}$. The quartic term in the superpotential is the simplest addition that results in a first-order, thermal  phase transition in the gauge theory (for appropriate values of $\phi_M$ and $\phi_Q$). The sextic term in the superpotential  guarantees that the five-dimensional geometry is regular even in the zero-temperature limit. Finally, for the chosen values \eqn{param} of the parameters, we will see in the next section that the phase diagram is generic in the sense that all quantities are of order unity in units of $\Lambda$. 

For completeness, we note that varying the values of $\phi_M$ and $\phi_Q$ changes the thermodynamic properties of the dual gauge theory. For example, keeping $\phi_Q=10$ but making $\phi_M> 1.08$ turns the first-order transition into a smooth crossover. For an extensive discussion we refer refer the reader to \cite{Bea:2018whf}. 

\subsection{Gauge theory thermodynamics}
In the gauge theory we use Cartesian coordinates $\{t, x_i\}=\{t,x,y,z\}$ and work with energy density $\mathcal{E}$ and pressures 
$\mathcal{P}_i$ obtained by rescaling the stress tensor according to
\be
\label{pre1}
{\cal E} = - \frac{\kappa_5^2}{2\ell^3} \, \langle T^t_t \rangle \sac 
\mathcal{P}_i =\frac{\kappa_5^2}{2\ell^3} \,\langle T^i_i  \rangle\,,
\ee
with no sum over $i$ on the right-hand side. In thermal equilibrium all pressures are equal, $\mathcal{P}_i=\mathcal{P}$, and the free energy density is $\mathcal{F}=-\mathcal{P}$. For an $SU(\nc)$ gauge theory the prefactor on the right-hand side of \eqn{pre1} typically scales as $\nc^{-2}$, whereas the stress tensor scales as $\nc^2$. The rescaled quantities are therefore finite in the large-$N$ limit. The stress tensor and the expectation of the scalar operator are related through the Ward identity
\be
\label{ward}
\langle T^\mu_\mu \rangle = -\Lambda \langle \mathcal{O} \rangle \,.
\ee
In treatments of the Electroweak transition it is common to separate the scalar degree of freedom associated to the  expectation value of the Higgs field from the plasma degrees of freedom. In our treatment this is neither necessary nor natural, since the holographic stress tensor that we will determine in \sect{time} includes the contribution of all degrees of freedom in the theory.

\begin{figure}[th]
\begin{center}
\includegraphics[width=.8\textwidth]{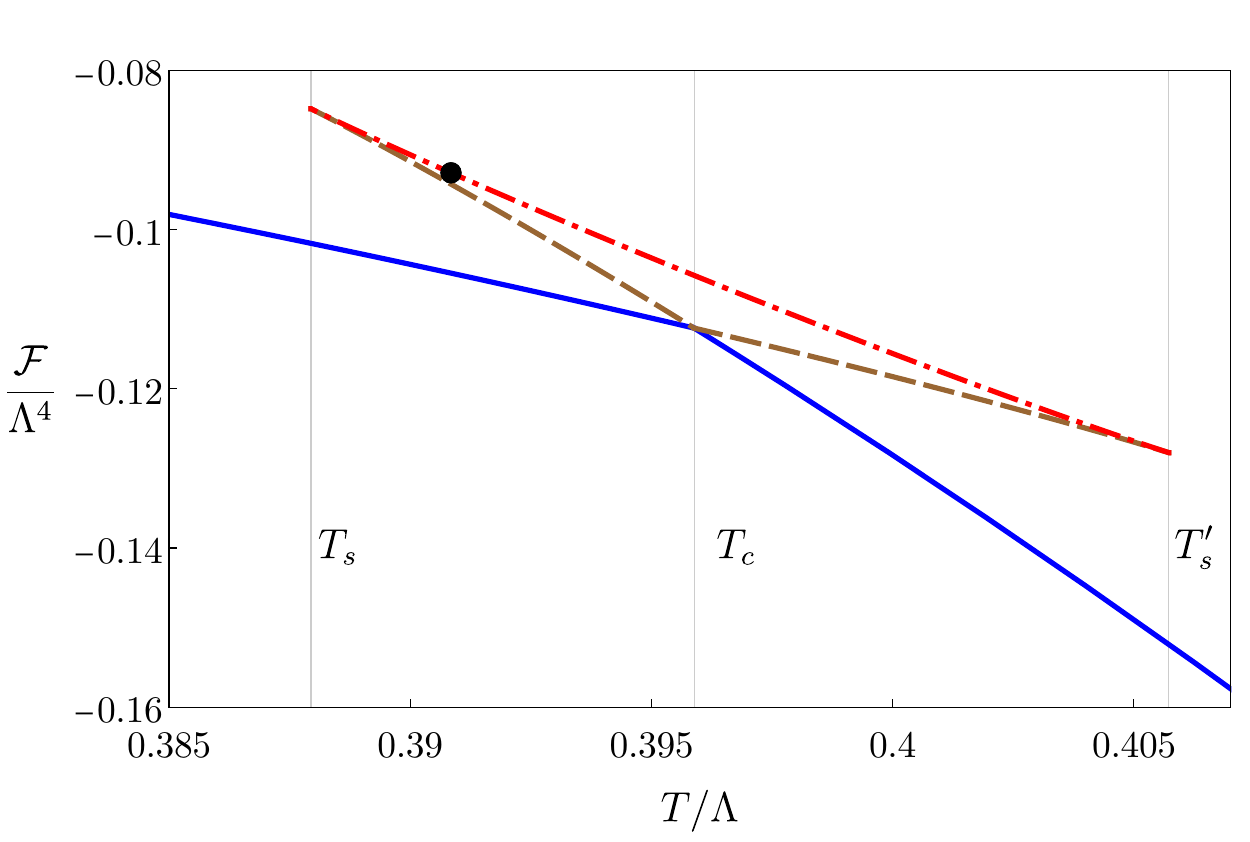} \\[2mm]
\includegraphics[width=.8\textwidth]{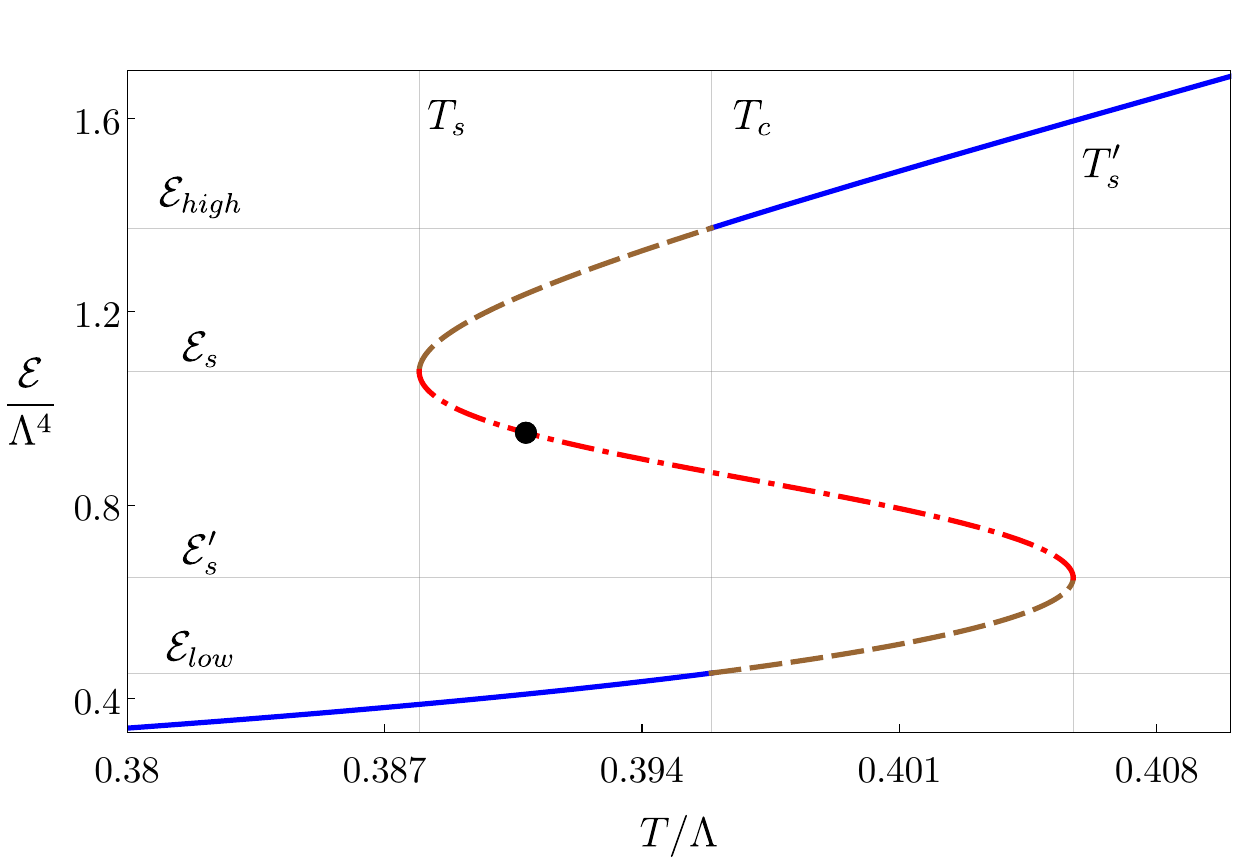}
\end{center}
\caption{Free energy density (top) and energy density (bottom) of the four-dimensional gauge theory dual to \eqn{action}-\eqn{param}. States on the solid, blue curves are thermodynamically stable. States on the dashed, brown curves are metastable. States on the dashed-dotted, red curve are unstable. The black circle with $T=0.3908\Lambda$ indicates the initial state on which we will focus in this paper. }
\label{free}
\end{figure}

The thermodynamics of the gauge theory can be extracted from the homogeneous black brane solutions on the gravity side (see e.g.~\cite{Gubser:2008ny}).
\fig{free} shows the results for the free energy density (top) and the energy density (bottom) as a function of  temperature. 
In \fig{free}(top) we see the familiar swallow-tail behaviour characteristic of the free energy for a first-order phase transition. In \fig{free}(bottom)  this leads to the usual multivaluedness of the energy density as a function of temperature. At high and low temperatures there is only one phase available to the system. Each of these phases is represented by a solid, blue curve. In \fig{free}(top) these two curves cross at a critical temperature 
\be
\label{tc}
T_c=0.396\Lambda \,. 
\ee
At this point the state that minimizes the free energy  moves from one branch to the other. The first-order nature of the transition is encoded in the non-zero latent heat, namely in the discontinuous jump in the energy density in \fig{free}(bottom) given by 
\be
\label{lat}
\mathcal{E}_\textrm{latent}=\E_\textrm{high}-\E_\textrm{low} =0.92 \, \Lambda^4\,.
\ee
Note that the phase transition is a transition between two deconfined, plasma phases, since both phases have energy densities of order $N^2$ and they are both represented by a black brane geometry with a horizon on the gravity side. 

In a region 
\be
\label{ts}
\Ts = 0.3879\Lambda < T < \Ts'=0.4057\Lambda
\ee
 around the critical temperature  there are three different states available to the system for a given temperature. The thermodynamically preferred one is the state that minimizes the free energy, namely a state on one of the blue curves. The states on the dashed, brown curves are not globally preferred but they are locally thermodynamically stable, i.e.~they are metastable. This follows from the convexity of the free energy, which indicates a positive specific heat
\be
\label{heat}
c_v \equiv \frac{d \mathcal{E}}{dT} \,.
\ee
This can be more clearly seen from the positive slope of the dashed, brown curves in \fig{free}(bottom). At the temperatures $\Ts$ and $\Ts'$ the metastable curves meet the dotted-dashed, red curve. States on the latter are locally unstable since their specific heat is negative. This region is known as the spinodal region. 

In terms of an effective potential for the expectation value of the scalar operator, 
$ \langle \mathcal{O} \rangle$, the region outside the range $(\Ts,\Ts')$ corresponds to temperatures for which  the potential has only one minimum and therefore there is a unique available state. In contrast, for  temperatures within the range $(\Ts,\Ts')$ the effective potential has two minima and one maximum. The global minimum corresponds to a stable state on a blue curve, the local but not global minimum corresponds to   a metastable state on a brown curve, and the maximum corresponds to an unstable state on the red curve. 
Plots of the effective potential, together with discussions of its properties, can be found in \cite{smith2023talk,Jorge,Mateos:toappear}.

As anticipated, we see that the phase diagram in \fig{free} is generic in the sense that there are no large ratios of energy densities, in contrast to e.g.~\cite{Attems:2017ezz,Attems:2019yqn}, where $\E_\textrm{high}$ and $\E_\textrm{low}$ differed by more than two orders of magnitude.

\subsection{Spinodal instability}
\label{spinodal}
\begin{figure}[t]
	\begin{center}
			\includegraphics[width=.7\textwidth]{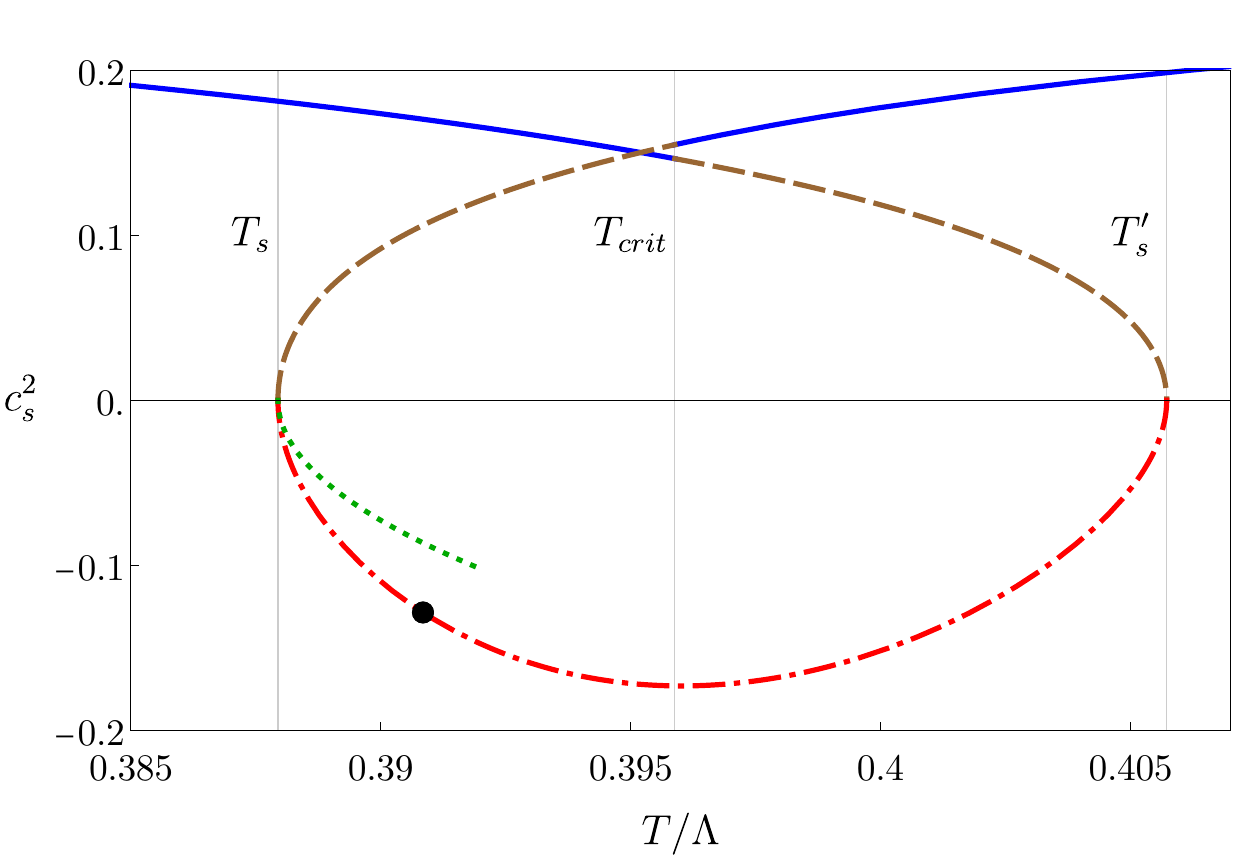}\\
			\includegraphics[width=.7\textwidth]{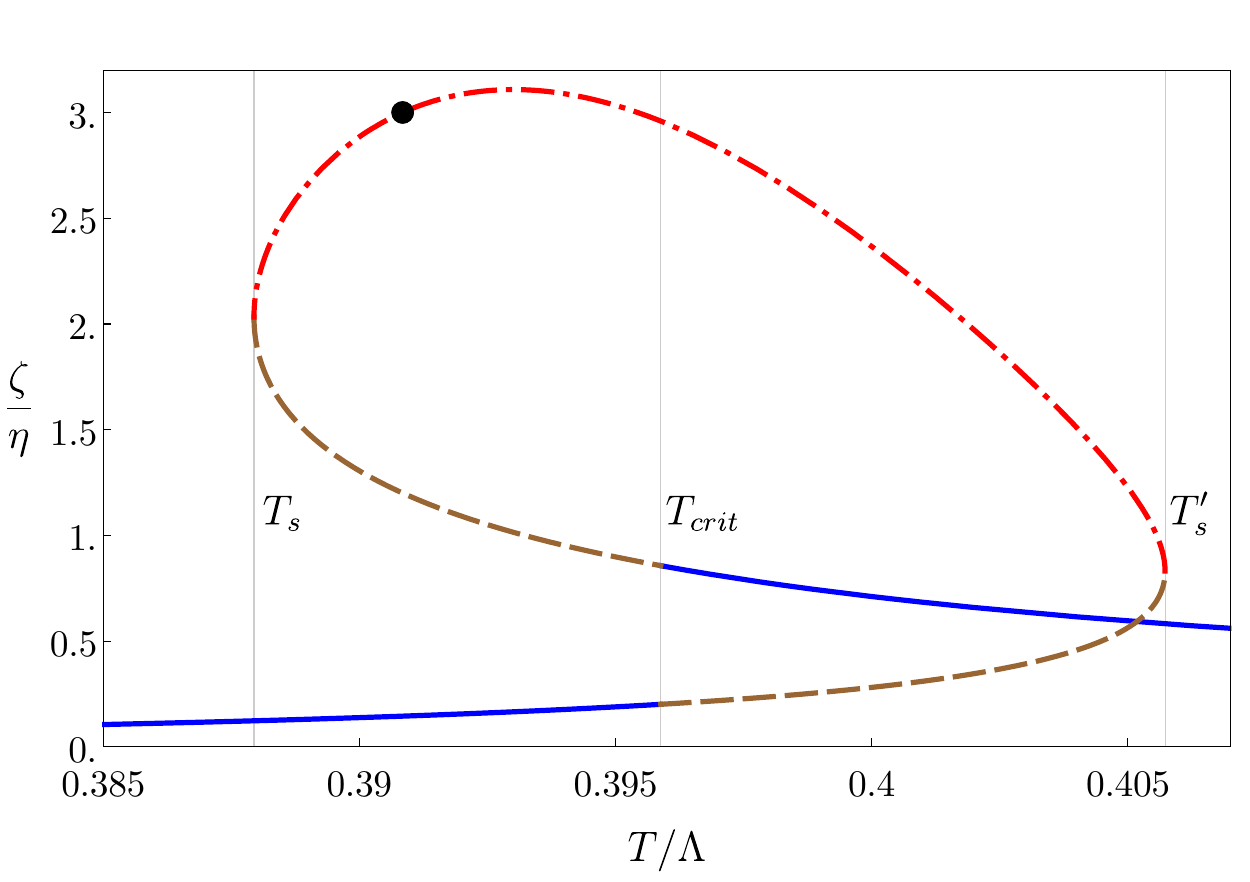}
		\caption{\label{cs2} 	 \small Speed of sound squared (top) and ratio of the bulk viscosity over the shear viscosity (bottom) versus temperature for the gauge theory dual to \eqn{action}-\eqn{param}. The color coding is as in Fig. \ref{free}. The dotted, green curve in the top figure corresponds to the approximation \eqn{csclose}.
		}
	\end{center}
\end{figure}

States on the dashed-dotted red curves of \fig{free} are locally thermodynamically unstable since the specific heat  is negative, $c_v<0$.  These states are also dynamically unstable. The connection with the dynamic instability was pointed out in an analogous context in \cite{Buchel:2005nt} and it arises as follows. The speed of sound is related to the specific heat $c_v$ and the entropy density $s$ through 
\be
\label{cs}
c_s^2 \equiv \frac{d\mathcal{P}}{d \mathcal{E}} =  
\left( \frac{d\mathcal{P}}{d T} \right) \left( \frac{d\mathcal{E}}{d T}\right)^{-1} = \frac{s}{c_v} \,.
\ee
Since the entropy density is positive everywhere,  $c_s^2$ is negative on the dashed-dotted, red curves of \fig{free}, as shown in \fig{cs2}(top), and consequently $c_s$ is purely imaginary. 
The amplitude of long-wave length, small-amplitude sound modes behaves as 
\be
\mathcal{A} \sim e^{-i \omega(k) t } \, , 
\label{amplitude0}
\ee
with a dispersion relation given by (see e.g.~Appendix A of \cite{Attems:2019yqn} for the derivation)
\be
\label{disp}
\omega_{\pm} (k) =\pm  \, c_s k  - \frac{i}{2} \Gamma k^2 + O(k^3) \,.
\ee 
We emphasize that, as indicated, this expression is an expansion valid at low momentum. In other words, it is a hydrodynamic approximation to the exact dispersion relation; we will come back to this point in \Sec{time}. 
The two signs give rise to one stable mode and one unstable mode. The sound attenuation constant is given by 
\be
\label{ate}
\Gamma = \frac{1}{T} \left(\frac{4}{3} \frac{\eta}{s}  + \frac{\zeta}{s} \right)   
\ee
where  $\eta$ and $\zeta$  are the shear and  bulk viscosities, respectively. In our model $\eta/s=1/4\pi$ \cite{Kovtun:2004de}. We compute $\zeta$ numerically following \cite{Eling:2011ms} and we show the result  in \fig{cs2}(bottom).

\begin{figure}[t]
\begin{center}
\includegraphics[width=.8\textwidth]{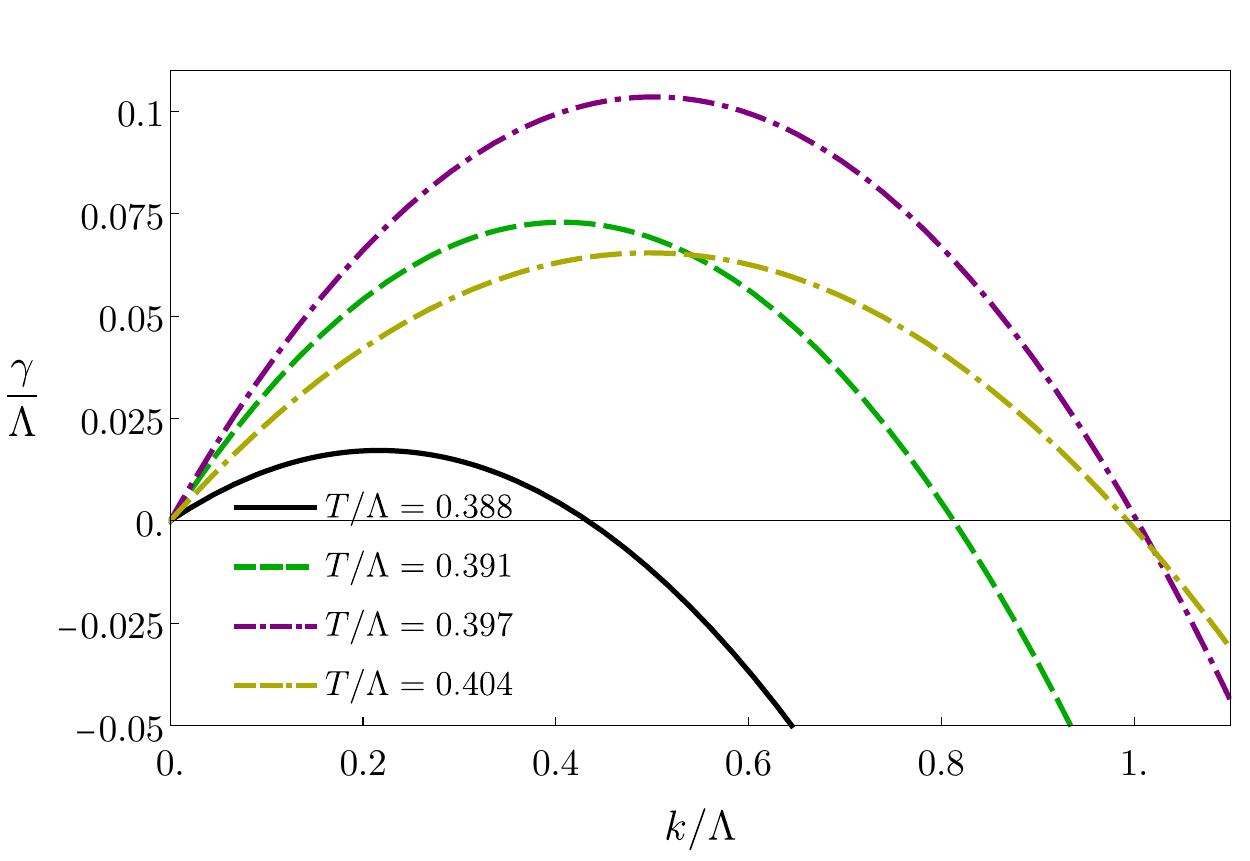}
\caption{\label{gamma}  \small
		Hydrodynamic approximation to the growth rates $\gamma(k)$ for different states on the spinodal region of  \fig{free}. 
}
\end{center}
\end{figure}

An imaginary value of $c_s$ leads to a purely real value of the growth rate
\be
\label{exponents}
\gamma(k) \equiv -i \omega (k) \,. 
\ee
For small momenta \eqq{disp} yields for the unstable mode
\be
\label{small}
\gamma (k) =  \left |c_s \right | k - \frac{1}{2} \Gamma k^2 +  O(k^3)\,.
\ee
We see that, in this hydrodynamic approximation, the dispersion relation gives rise to the familiar parabolas displayed in \fig{gamma}. 
The slope of these curves at $k=0$ is $\left |c_s \right |$. The non-monotonic behaviour of this quantity with the temperature observed in \fig{gamma} agrees with that on the red curve in \fig{cs2}(top).

The parabolas in  \fig{gamma}  depend on the energy density $\mathcal{E}$ of the state under consideration because both $c_s$ and $\Gamma$ depend on $\mathcal{E}$.
\begin{figure}[t!!]
	\begin{center}
			\includegraphics[width=.7\textwidth]{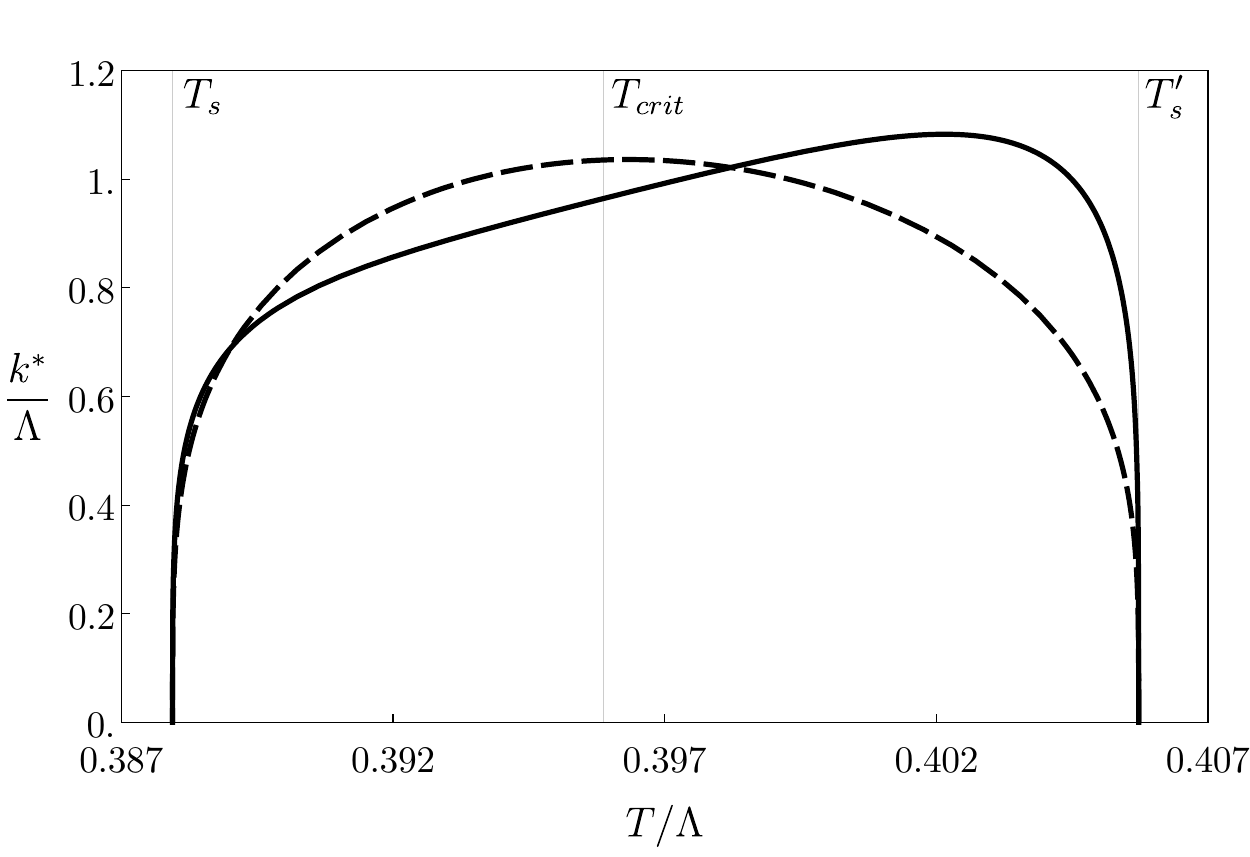}\\
			\includegraphics[width=.72\textwidth]{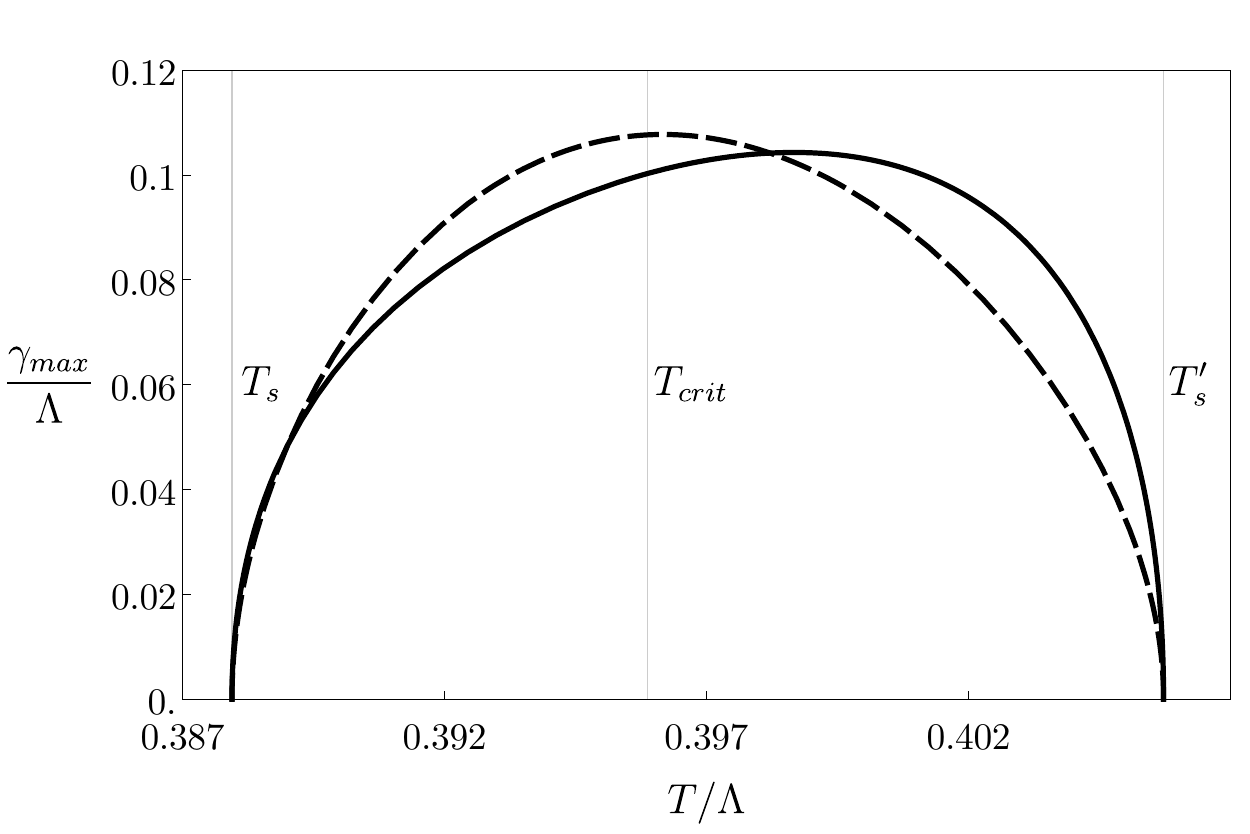}\,\,\,\,\,\,\,\,\,\,
		\caption{\label{kandgamma} 	 \small Comparison between the hydrodynamic values of $k_*$ (top) and $\gamma_\textrm{max}$ (bottom) given by \eqn{small} (solid curves) and the ball-park estimates \eqn{ball} (dashed curves).
		}
	\end{center}
\end{figure}
Working to order $k^2$ we see that the growth rate is positive for momenta in the range 
$0 < k < k_{*}$ with 
\be
\kstarhyd \simeq \frac{2\left | c_s \right |}{\Gamma} \,,
\ee
where the superscript is to emphasize that this is the hydrodynamic approximation to the exact value. We see that the spinodal instability is an infrared instability, since only modes with momentum below a certain threshold are unstable. The most unstable mode, namely the mode with the largest growth rate, has momentum and growth  rate given by
\be
\label{why}
\kmaxhyd \simeq \frac{\left | c_s \right |}{\Gamma} \sac 
\gmaxhyd \simeq \frac{\left | c_s \right |^2}{2 \Gamma} \,.
 \ee
 Note that 
\be
\gmaxhyd=\frac{1}{2} \left | c_s \right | \kmaxhyd\,.
\ee

A ballpark estimate for these quantities can be obtained by using $\eta/s=1/4\pi$ and 
approximating $\zeta/\eta \sim 2$ in the spinodal region, which implies 
\be
\Gamma \sim \frac{1}{\pi T} \,.
\ee
Under these conditions 
\be
\label{ball}
\kstarhyd \sim 2\left | c_s \right | \pi T \sac 
\kmaxhyd \sim \left | c_s \right | \pi T \sac 
\gmaxhyd \sim \frac{1}{2} \left | c_s \right |^2 \pi T \,.
\ee
As mentioned above, these quantities depend on  the energy density $\mathcal{E}$ of the state under consideration. In particular, they all go to zero as the state on the spinodal branch approaches one of the turning points at $T=\Ts, \Ts'$, since at those points $c_s=0$. As the state moves away from these points towards the interior of the spinodal region these quantities increase and reach a maximum. This is illustrated in \fig{kandgamma}, where we compare the values of $k_*$ and $\gamma_\textrm{max}$ given by \eqq{small} with  those given by the estimates  \eqn{ball}. We see that the latter provide a reasonable approximation in the entire range, and that this is better close to the turning points, which will be a particularly  important region below. The maxima of, and the  points where, the curves in \fig{gamma} cross the horizontal axis are consistent with the values of $k_*$ and $\gamma_\textrm{max}$ in \fig{kandgamma}. Rotating \fig{free} ninety degrees we see that, close to the turning point at $\Ts$, the temperature has a minimum as a function of the energy density. This means that, close to this point, 
\be
\label{two}
T-\Ts \sim \left( \mathcal{E} - \Es \right)^2 \,.
\ee
The exponent on the right-hand side can be understood as follows. In our model, the different equilibrium states in the phase diagram are in one-to-one correspondence with the value of the scalar field at the horizon, $\phi_\textrm{H}$. The energy density is a monotonically decreasing function of $\phi_\textrm{H}$, whereas the temperature has a minimum at the value of 
$\phi_\textrm{H}$ corresponding to $\Ts$. Expanding around this point and eliminating $\phi_\textrm{H}$ leads to \eqn{two}.\footnote{In our case $d^2T/d\phi_\textrm{H}^2\neq 0$ at $\Ts$. In cases in which this derivative vanishes, the exponent on the right-hand side of \eqref{two} would change.} Inverting this relation, computing the specific heat \eqn{heat}, and substituting in \eqn{cs} we find that, close to $\Ts$, 
\be
\label{csclose}
c_s^2 \simeq \frac{1}{\sqrt{\Ts}} \sqrt{T-\Ts} + \cdots \,,
\ee
where we have estimated the prefactor based on dimensional analysis. This approximation to the function $c_s^2(T)$ is shown as a green, dashed curve in 
\fig{cs2}(top).

We emphasize again that the analysis in this section is based on a hydrodynamic approximation to the dispersion relation. In \Sec{time} we will determine the exact dispersion relation and, in particular, the exact values of $k_*, \kmax$ and $\gmax$. For concreteness, in several places in the paper we will use the hydrodynamic dispersion relation $\gamma(k)$ since it has the advantage that it is analytically known. Nevertheless, most final results will only depend on $\kstarhyd, \kmaxhyd, \gmaxhyd$ and $|c_s|$. In these cases one can obtain the exact result  by simply replacing 
\be
\label{reprep}
\kstarhyd \to k_* \sac \kmaxhyd \to \kmax \sac 
\gmaxhyd \to \gmax \sac |c_s| \to \frac{2\gmax}{\kmax} \,.
\ee

To summarize this section, states on the  dashed, red curves of \fig{free} are afflicted by a dynamical instability, known as spinodal instability, whereby long-wave length, small-amplitude perturbations in the sound channel grow exponentially in time.  The growth rate of the unstable modes is zero at the turning points and increases as the state moves deeper into the spinodal region.

\section{Dynamics of a first-order phase transition}
\label{thermo}
We will now discuss the conditions for the phase transition to take place via the spinodal instability. An important observation is that the expansion of the Universe may not be driven by the same sector that undergoes the phase transition. In particular, the temperatures in these two sectors may be vastly different.  This may happen if the phase transition takes place in a hidden sector. If the hidden sector couples weakly to the inflaton, then the energy density injected into this sector at the end of reheating,  and hence its temperature, may be vastly smaller than in the visible, SM  sector. Therefore in the following we will distinguish between the temperature $T$ of the sector undergoing the phase transition and the radiation temperature $\Trad$ driving the expansion of the Universe. If the phase transition takes place in the sector driving the expansion then these two temperatures coincide, otherwise they may differ.

\subsection{Suppressed bubble nucleation}
We  imagine that the sector where the phase transition will take place  starts off in the high-temperature, stable branch represented by the upper blue, solid curve in \fig{free}(bottom). As the Universe cools down the temperature falls below $T_c$ and we enter the metastable branch. Bubbles of the stable phase represented by the lower blue, solid curve in \fig{free}(bottom) can then begin to nucleate. This process requires thermal fluctuations large enough to decrease the energy density from the metastable phase to the stable phase in a finite region of space with a volume equal to that of the critical bubble. The larger the number of degrees of freedom that participate in the transition the more unlikely these fluctuations are, the more suppressed the bubble nucleation rate becomes, and the closer to $\Ts$ bubbles begin to be effectively nucleated. Under these circumstances  there are two crucial aspects that are often  overlooked in usual treatments. First, the nucleated bubbles are very small both in size and in amplitude and hence they need time to grow to the point where the energy density in their interior reaches that of the stable phase. Second,  there is a finite available time for the transition. Under these conditions the  arguments usually employed to compute the duration of the transition are misleading. We will therefore repeat the standard calculation paying attention to these two aspects. In order to be concrete about the large number of degrees of freedom involved in the transition we will assume that the theory in question is a large-$N$ gauge theory. We will see that if $N$ is sufficiently large then the transition cannot be completed via bubble nucleation. Instead, the temperature eventually reaches $\Ts$ and the Universe enters the spinodal region.    Recent analyses of bubble dynamics in large-$N$ gauge theories include \cite{Ares:2021ntv,Ares:2021nap}.

We follow the discussion in Sec.~7.1 of \cite{Hindmarsh:2020hop}. The starting point is the expression for the fraction of the volume of the Universe that remains in the metastable phase of the sector undergoing the phase transition. Let $t_c$  the time at which the temperature of this sector crosses $T_c$. Then at a time $t > t_c$ this fraction is 
\be
\label{h}
h(t) = e^{-I(t)}  \,,
\ee
with 
\be
\label{i}
I(t)=\int_{t_c}^t dt' \, \frac{4\pi}{3} v^3 (t-t')^3 \, \frac{\Gamma(t')}{V}\,.
\ee
In this expression $v$ is the bubble wall terminal velocity. The integrand is the volume of a bubble nucleated at time $t'<t$ multiplied by the bubble nucleation rate per unit volume, which takes the form 
\be
\label{Gamma}
\frac{\Gamma (T)}{V} \sim 
T^4 \, e^{-S(T)} \,.
\ee
Here $S(T)$ is the action of the instanton that mediates the transition, namely the critical bubble, and $\Gamma (T)$ should not be confused with the sound attenuation constant \eqn{ate}.  
 Eqn.~(\ref{i}) assumes that the interior of a nucleated bubble is instantaneously in the stable phase, and that the bubble wall instantaneously begins to move with the corresponding terminal velocity $v$. We will come back to this below.

Let $\ts$ be the time at  which the temperature in the sector undergoing the phase transition reaches $\Ts$. Below we will be interested in evaluating the fraction $h(t)$ at the latest possible time, namely at $t=\ts$. A crucial point is that the instanton action must vanish at the turning point, since there the barrier to nucleate bubbles disappears. Therefore, 
close to this point we have
\be
\label{S}
S(T) \sim N^2 \, \left( \frac{T-\Ts}{\Ts} \right)^x \,,
\ee
where $x$ is a constant and we have explicitly displayed the $N^2$ scaling of the action expected for a large-$N$ gauge theory. Studies based on effective actions for the order parameter with a canonically normalized kinetic term yield  $x = 3/2$ \cite{Enqvist:1991xw}, and we will assume this as a ballpark value in our discussion at the end of this section. However, in generic theories the kinetic term is not canonically normalized, see e.g.~\cite{Ares:2021nap,Ares:2021ntv,Henriksson:2025vci,Mateos:toappear} for holographic examples. Although this modifies the value of $x$, this value is still of order unity \cite{Mateos:toappear}.   

Unless $T$ is very close to $\Ts$, the $N^2$-scaling in \eqn{S} makes the instanton action very large, which in turn makes $\Gamma$ and $I$ exponentially close to zero, which makes $h$ exponentially close to one. Let us define the nucleation temperature, $\Tn$,  as the temperature at which this suppression disappears, namely the temperature at which 
\be
S(\Tn) \sim 1\,.
\ee
This is given by
\be
\label{TH}
\Tn = \Ts + \frac{\Ts}{N^{2/x}} \,.
\ee
Up to logarithmic corrections, this is also the temperature at which  one bubble per  Hubble time per Hubble volume is nucleated, namely the temperature such that 
\be
\frac{\Gamma (\Tn)}{V} \sim H^4 \,.
\ee
As the Universe cools down, the temperature drops below $T_c$, but the phase transition does not start until the time $\tn$ at which the temperature reaches $\Tn< T_c$. The remaining available time for the phase transition to take place is therefore 
\be
\label{delta}
\Delta t \,= \, \ts-\tn = \frac{1}{H \Ts}\,
 (\Tn-\Ts) \sim \frac{1}{H N^{2/x}} \,,
\ee
where we have used the usual relation to translate between time and temperature
\be
\label{ht}
\frac{dt}{dT} = -\frac{1}{T H }  \,.
\ee
Note that the Hubble rate $H$ is controlled by $\Trad$ through the Friedmann equation
\be
\label{H}
H^2\sim G \Trad^4 \sim \frac{\Trad^4}{M_p^2} \,,
\ee
where $G$ is the four-dimensional Newton's constant and $M_p$ is the (reduced) Planck mass. Recall that $\Trad$ may or may not coincide with $T$. 

The key point is that, because $\Tn$ is parametrically close to $\Ts$, the bubbles that get nucleated are small bubbles that do not have the stable phase at their center. On the contrary, as $\Tn$ approaches $\Ts$, the energy density at the centre of the nucleated bubbles approaches the energy density of the metastable phase \cite{Enqvist:1991xw}. These bubbles need a certain time to grow in size and in amplitude until the energy density at their centre reaches that of the stable phase. If this time is too long compared to the available time \eqn{delta} then no volume will be occupied by the stable phase when the system reaches the spinodal region. Since the critical bubbles are unstable, we expect their growth to be of the form $\exp(\gamma t)$. A natural estimate for the growth rate is $\gamma \sim \Ts \sim T_c$,  since we expect this to be set by a microscopic scale. Comparing the growth time $\gamma^{-1}$ to $\Delta t$ we see that the condition that the transition does not take place, $\gamma^{-1} \gtrsim \Delta t$, translates into 
\be
\label{n1}
N^{2/x} \gtrsim  \frac{T_c}{H}  \sim \frac{T_c \, M_p}{\Trad^2} \,,
\ee
where in the second equation we have used \eqn{H}.

Before we analyse the condition \eqn{n1}, we note that, if it is not satisfied, then the transition does take place via bubble nucleation. In this case the growth time of the bubbles is short compared to $\Delta t$ and we can neglect it. We can  also neglect the time it takes the bubbles to accelerate to their terminal velocity $v$, since presumably this time is also of order $1/T_c$. We can then evaluate what fraction of the volume has been converted to the stable phase between the times $\tn$ and $\ts$. For this purpose we use \eqn{Gamma} and \eqn{S} in \eqn{i} and we change the integration variable from time to temperature according to 
\be
\ts-t' = \frac{1}{H \Ts}\, (T' - \Ts) \,.
\ee
The result is 
\be
I(\Ts) = \left( \frac{4\pi}{3} v^3 \right) \left( \frac{\Ts}{\Hs} \right)^4   
\int_{\Ts}^{\Tn}\,  \frac{dT'}{\Ts} \left( \frac{T'-\Ts}{\Ts} \right)^3\, 
\exp \left[ -N^2 {\left(  \frac{T'-\Ts}{\Ts} \right)}^x \,\right] \,.
\ee
Because of the exponential suppression we can extend the integral to $\Tn \to\infty$. Changing variables through 
\be
y= \frac{T'-\Ts}{\Ts} 
\ee
we then get 
\be
I(\Ts) = \left( \frac{4\pi}{3} v^3 \right) \left( \frac{\Ts}{\Hs} \right)^4   
\int_{0}^{\infty} dy \, y^3 \exp \left(- N^2 y^x \right) \,.  
\ee
The integral can be performed explicitly and we arrive at 
\be
I(\Ts) = \left( \frac{4\pi}{3} v^3 \right) \left( \frac{\Ts}{\Hs} \right)^4  
\, 
\frac{1}{N^{8/x}} \, \frac{\Gamma(4/x)}{x} \,.
\ee
Focussing on the parametric dependence we see that the condition for $I(\Ts)\lesssim 1$ is the same as (\ref{n1}). 

Eqn.~(\ref{n1}) shows that the transition via bubble nucleation can always be prevented if $N$ is sufficiently large. This is simply a consequence of the fact that the time available for the transition decreases parametrically with $N$ and that the nucleated bubbles near $\Ts$ need a finite amount of time to grow to the stable phase. To estimate the value of $N$ we must distinguish two cases. If the expansion of the Universe is driven by the same sector undergoing the phase transition then $\Trad=T_c$. This results in a fairly large value of $N$ unless $\Trad$ is close to $M_p$.  For example, for a phase transition at a GUT scale this ratio  is \mbox{$M_p/T_c \sim 10^2$}, which gives  $N\gtrsim 7$, whereas for a transition at the electroweak scale we have \mbox{$M_p/T_c \sim 10^{16}$}, resulting in $N\gtrsim 10^{6}$. Suppose instead that the transition takes place in a hidden sector at a temperature $T_c< \Trad$, as would be the case  if this sector couples to the inflaton more weakly than the SM degrees of freedom. In this case the condition to suppress bubble nucleation results in  mild constraints on the parameters of the hidden sector for two reasons. First, large values of $N$ are natural,  since string theory  compactifications can give rise to gauge-group ranks as large as  $O(10^5)$ \cite{Candelas:1997pq,Halverson:2015jua,Taylor:2015xtz,Taylor:2017yqr}. Second, even if $N$ is of order unity the condition \eqn{n1} only requires $T_c \lesssim \Trad^2/M_p$, which still results in a huge range of possible values for $T_c$.

\subsection{Thermal inflation}
When the Universe enters the metastable branch it is initially undergoing decelerated expansion, since it is radiation-dominated. We have seen that if $N$ is large enough then the Universe will supercool along the metastable branch until it enters the spinodal branch. Along this process the Universe may or may not enter a period of thermal inflation  \cite{Lyth:1995hj,Lyth:1995ka}, namely a period of accelerated expansion. Although in our model \eqn{action}-\eqn{superpotential} with the choice of parameters \eqn{param} this does not happen, we will now show that it would happen for other choices of the parameters. 

Whether thermal inflation takes place depends on the equation state of the model, specifically on the sign of the combination $\mathcal{E}+3\mathcal{P}$. If this  becomes negative along the metastable branch then the  acceleration of the Universe becomes positive and we enter a period of inflation. This can be equivalently characterized in terms of the ratio $w=\mathcal{P}/\mathcal{E}$. If $w<-1/3$ then the Universe inflates. Further, if $w=-1$ then the Universe inflates exponentially. 

\begin{figure}[h!!!]
\begin{center}
\includegraphics[width=.8\textwidth]{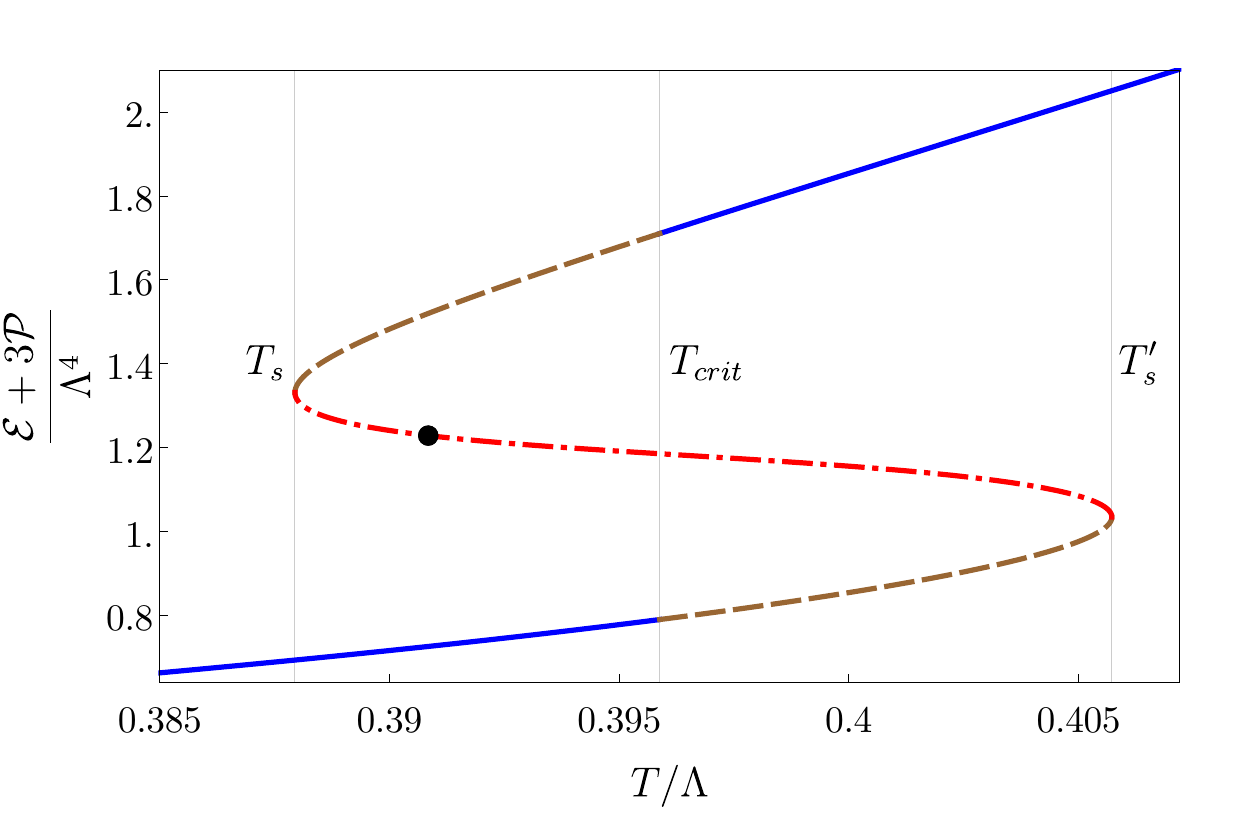}
\caption{\label{Eplus3P}  \small
		Energy density plus three times the pressure for the four-dimensional gauge theory dual to \eqn{action}-\eqn{superpotential} with the choice of parameters \eqn{param}. 
}
\end{center}
\end{figure}

\begin{figure}[h!!!]
	\begin{center}
			\includegraphics[width=.63\textwidth]{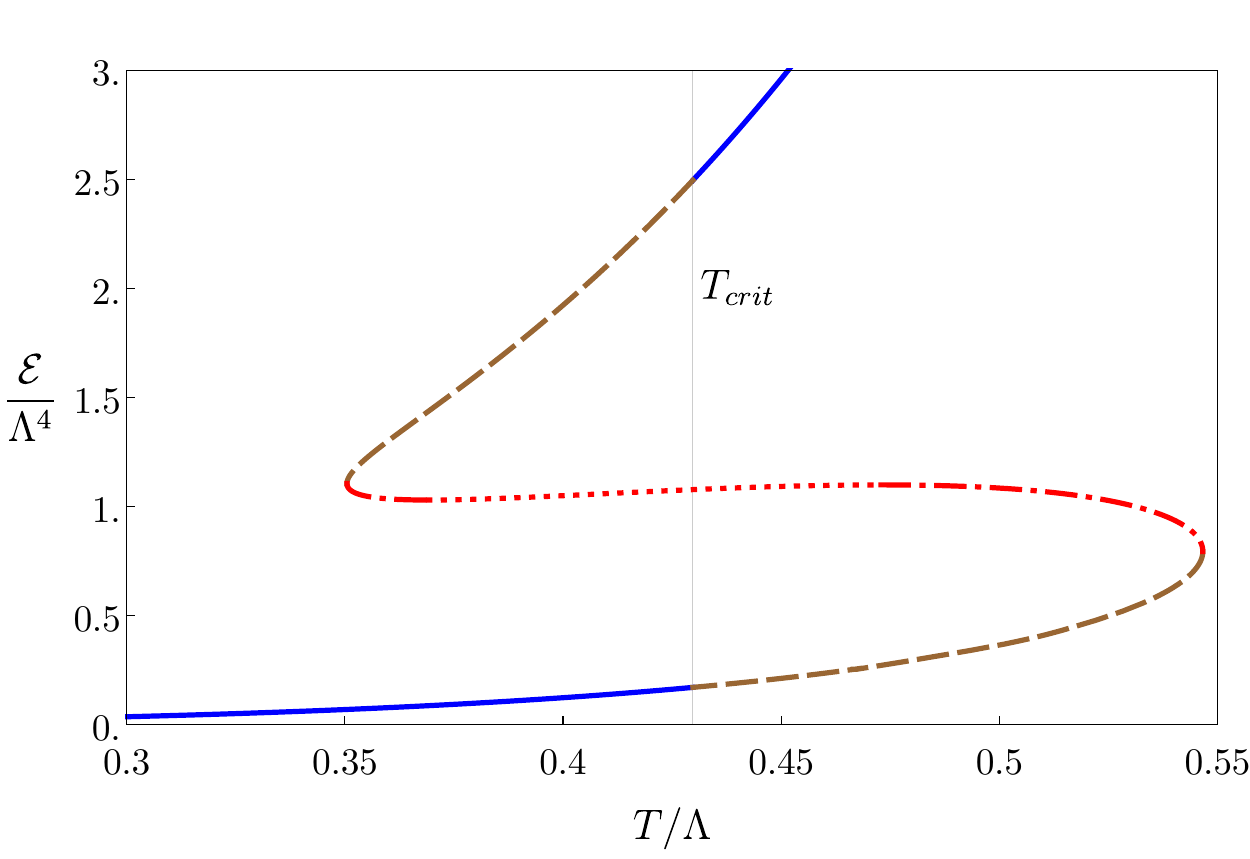}\\
			\includegraphics[width=.63\textwidth]{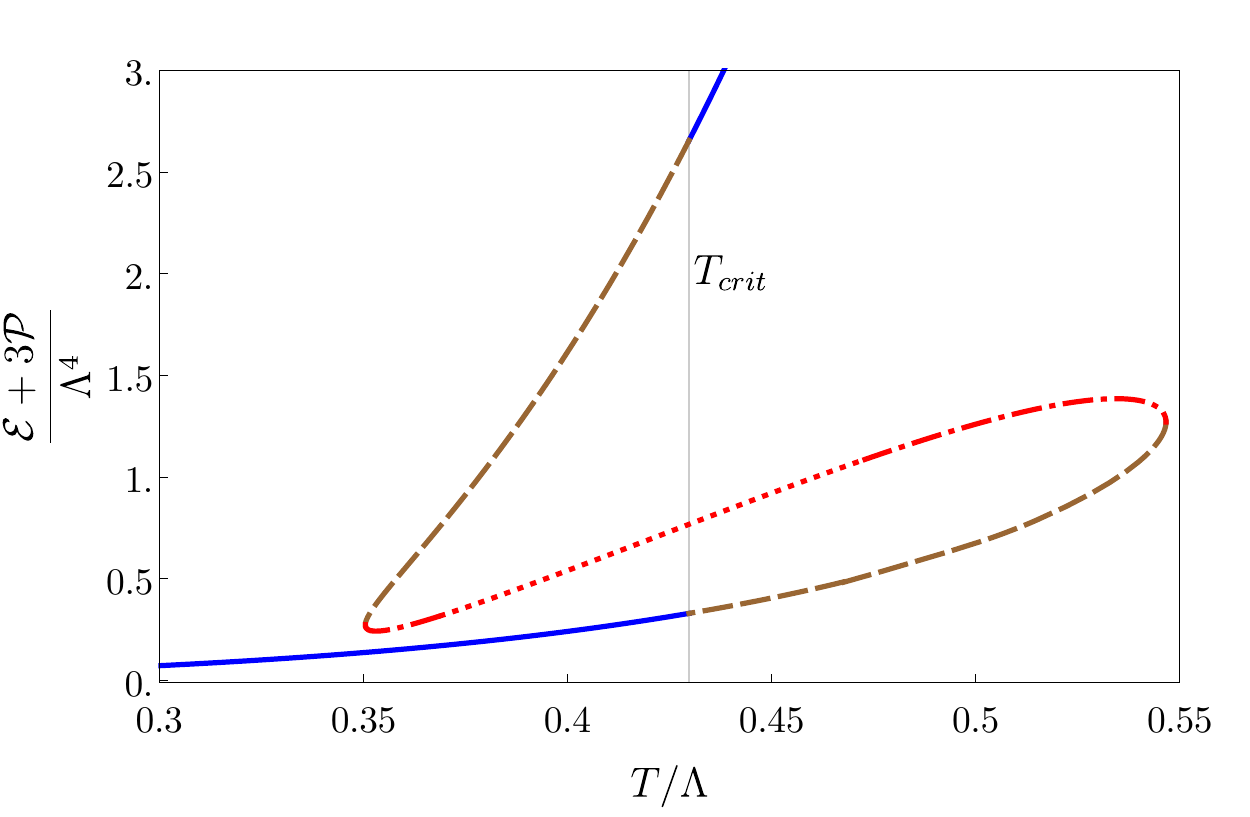}\\
			\includegraphics[width=.63\textwidth]{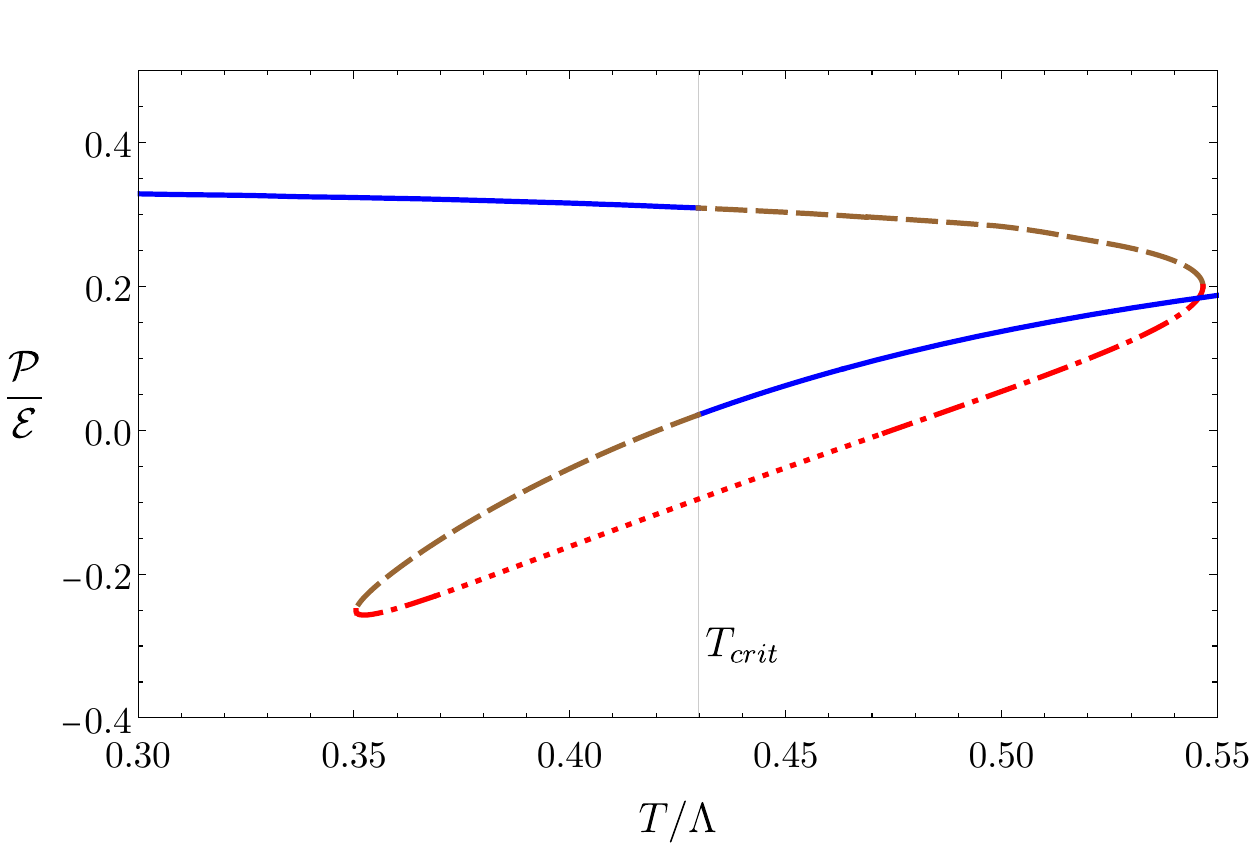}
		\caption{
		\label{theory3} 
		\small $\mathcal{E}$ (top), $\mathcal{E}+3\mathcal{P}$ (middle) and 
		$\mathcal{P}/\mathcal{E}$ (bottom) for the four-dimensional gauge theory 		dual to \eqn{action}-\eqn{superpotential} with the choice of parameters 			$\phi_M=0.8, \phi_Q = 10.$ (Plots reproduced from \cite{Bea:2018whf}.)
		}
	\end{center}
\end{figure}

\begin{figure}[h!!!]
	\begin{center}
			\includegraphics[width=.61\textwidth]{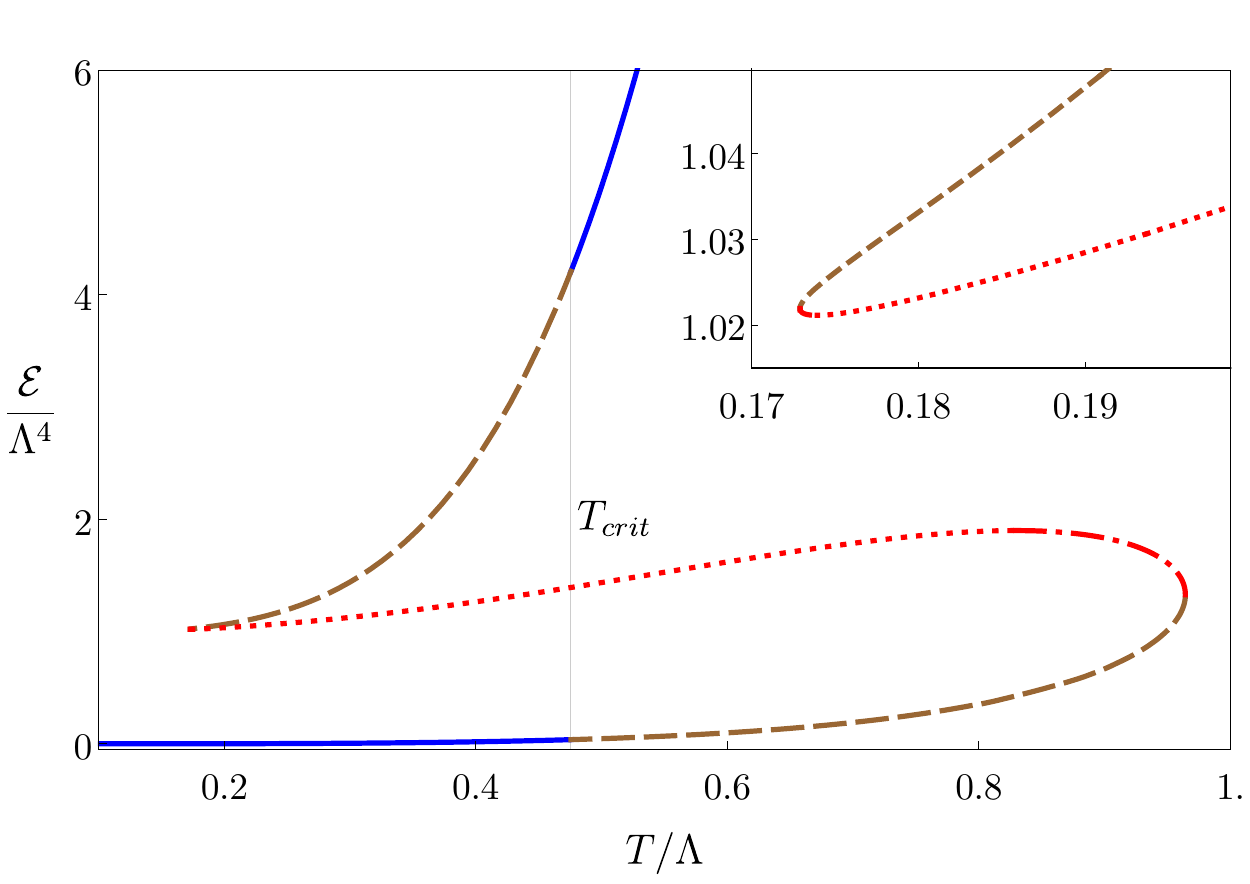}\\
			\includegraphics[width=.61\textwidth]{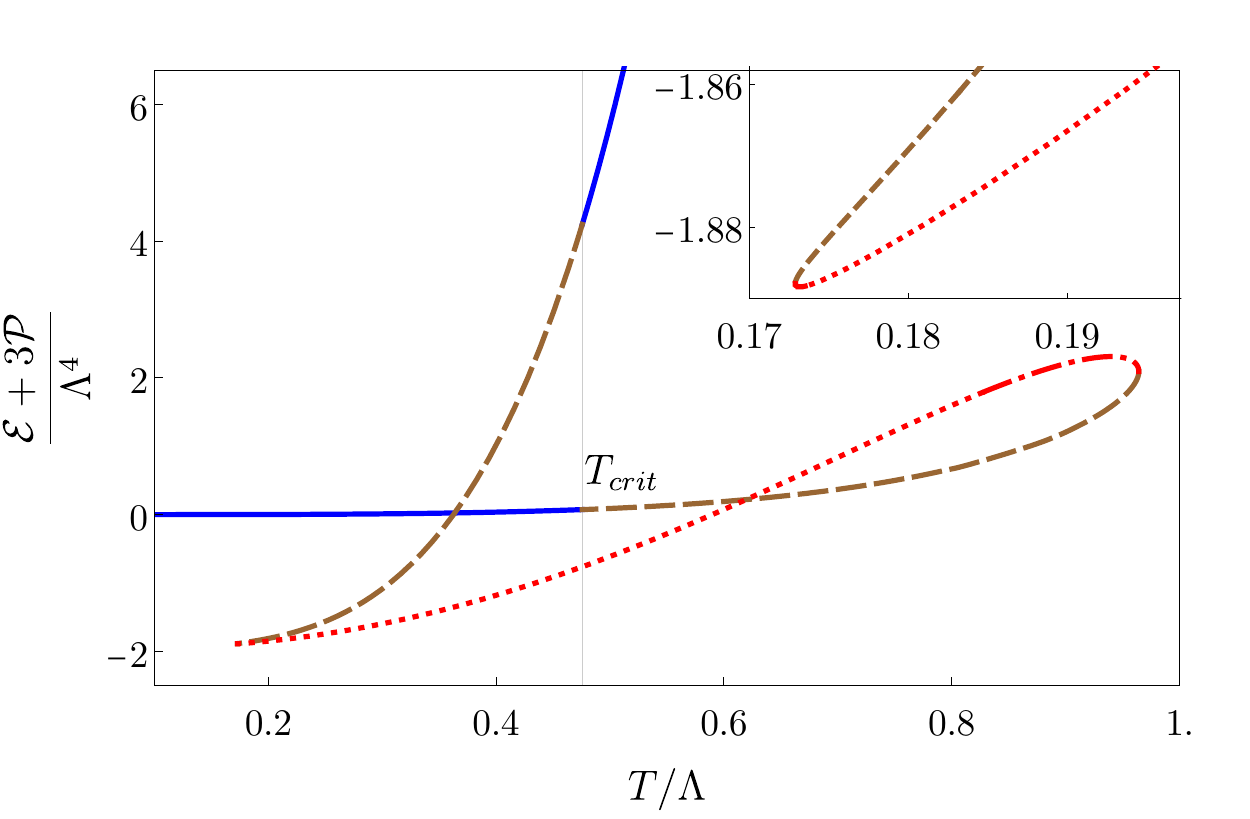}\\
			\includegraphics[width=.61\textwidth]{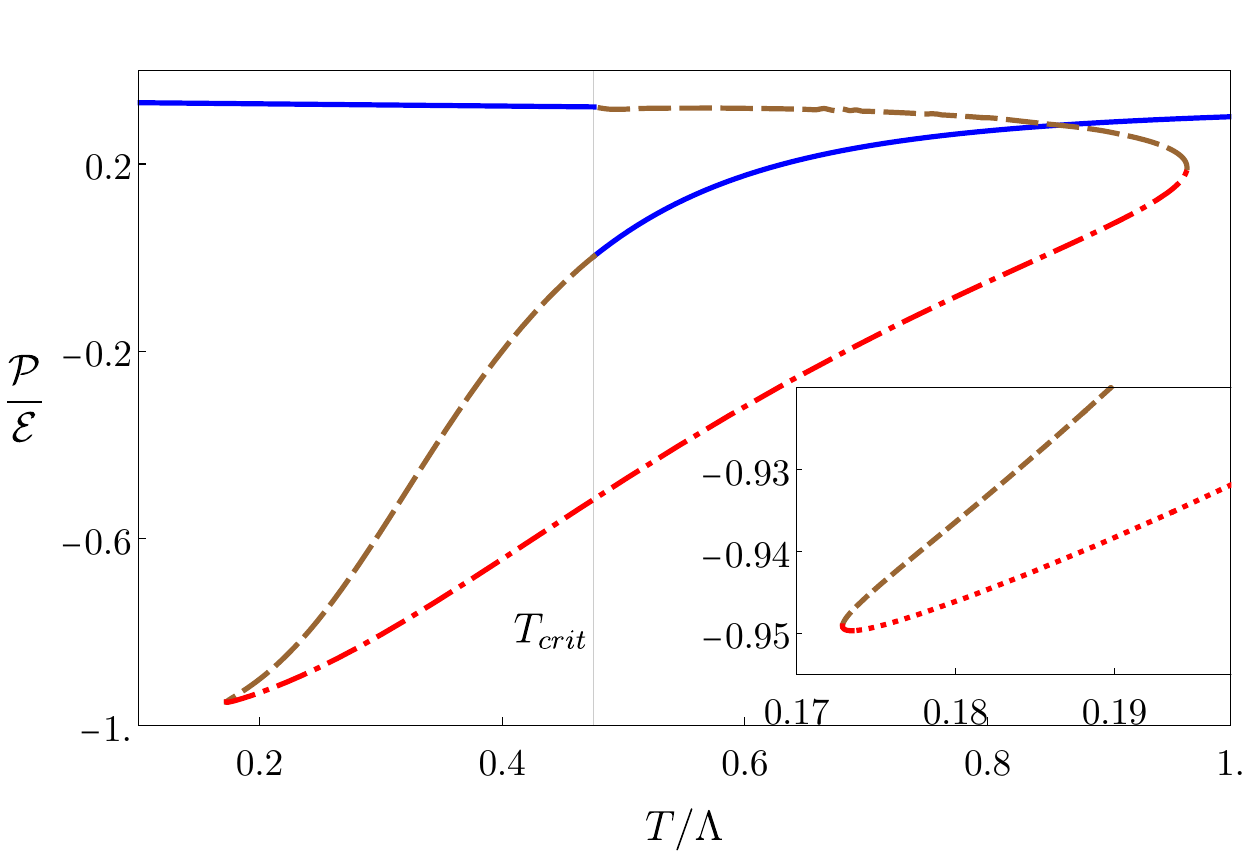}
		\caption{
		\label{theory2} 
		\small  $\mathcal{E}$ (top), $\mathcal{E}+3\mathcal{P}$ (middle) and 
		$\mathcal{P}/\mathcal{E}$ (bottom) for the four-dimensional gauge theory 		dual to \eqn{action}-\eqn{superpotential} with the choice of parameters 			$\phi_M=0.64, \phi_Q = 10.$ Note that the region close to $\Ts$ is smooth, not a cusp, as shown by the insets. (Plots reproduced from \cite{Bea:2018whf}.)
		}
	\end{center}
\end{figure}

\begin{figure}[h!!!]
	\begin{center}
			\includegraphics[width=.62\textwidth]{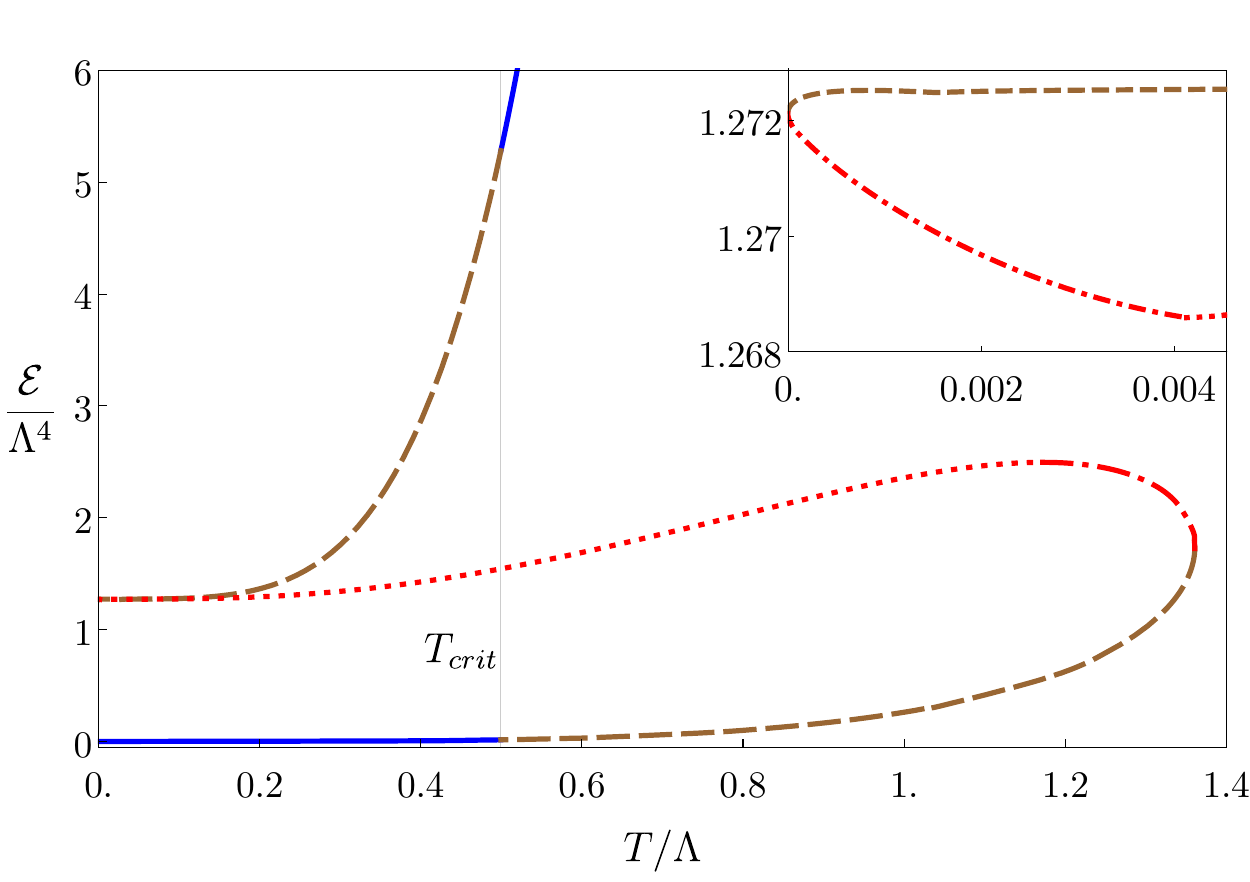}\\
			\includegraphics[width=.62\textwidth]{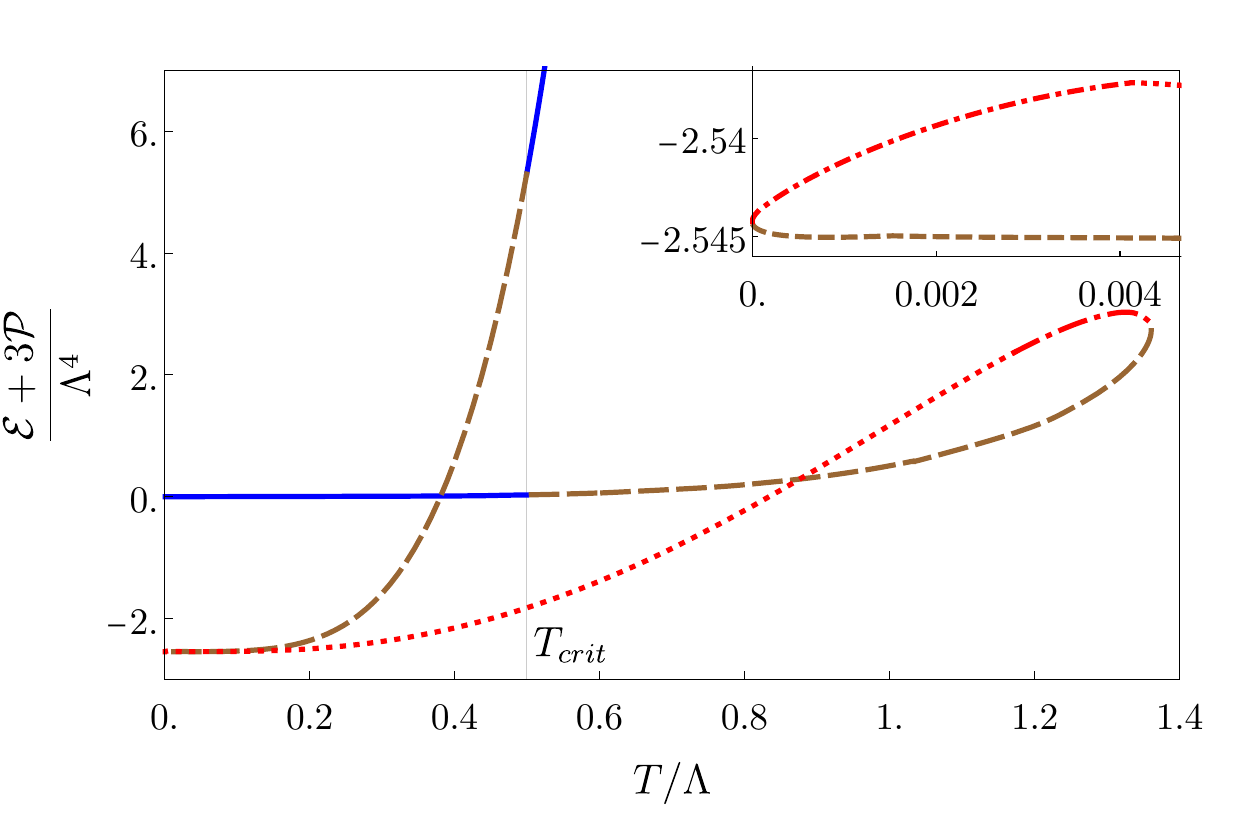}\\
			\includegraphics[width=.62\textwidth]{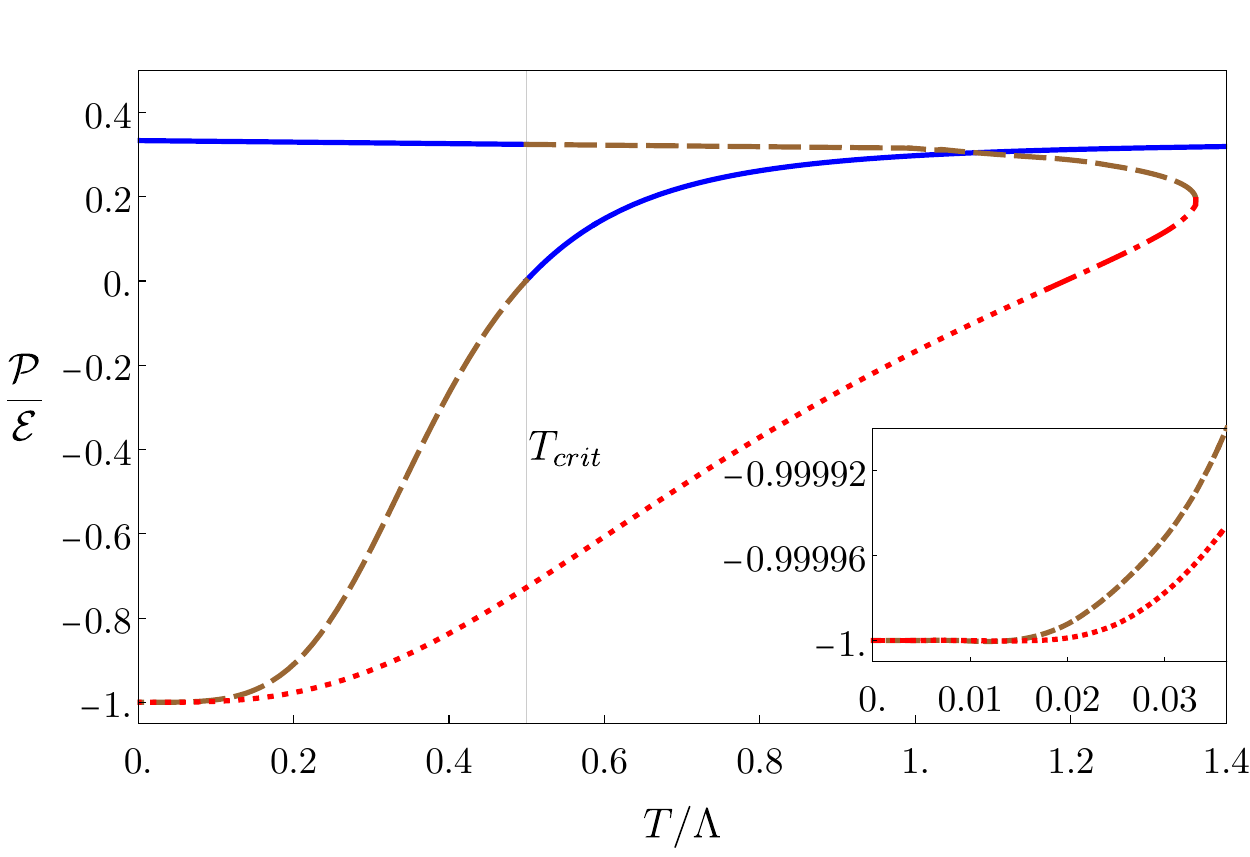}
		\caption{
		\label{theory4} 
		\small  $\mathcal{E}$ (top), $\mathcal{E}+3\mathcal{P}$ (middle) and 
		$\mathcal{P}/\mathcal{E}$ (bottom) for the four-dimensional gauge theory 		dual to \eqn{action}-\eqn{superpotential} with the choice of parameters 			$\phi_M=0.5797, \phi_Q = 10.$ 
		(Plots reproduced from \cite{Bea:2018whf}.)
		}
	\end{center}
\end{figure}

The combination $\mathcal{E}+3\mathcal{P}$  for the model \eqn{action}-\eqn{superpotential} with the choice of parameters \eqn{param} is shown in \fig{Eplus3P}. 
We see that it is positive everywhere and hence there is no period of thermal inflation.  Nevertheless, this same model can lead to thermal inflation for other choices of the parameters. In other words, by varying these parameters we can continuously interpolate between models with and without thermal inflation. To see this let us keep $\phi_Q=10$ fixed and vary $\phi_M$, as in \cite{Bea:2018whf}. The results for  $\phi_M=0.8$, $\phi_M=0.64$ and $\phi_M=0.5797$ are illustrated in  Figs.~\ref{theory3}, \ref{theory2} and \ref{theory4}, respectively. 

In \fig{theory3} we see that $\mathcal{E}+3\mathcal{P}$ attains a smaller value than in the case $\phi_M=1$, but it remains positive. Similarly, $w$ remains above $-1/3$. In contrast,  \fig{theory2} shows that $\mathcal{E}+3\mathcal{P}$ becomes negative in the lowest part of the upper metastable branch, which would lead to a period of thermal inflation. In fact, at the end of the metastable branch $w$ comes close to the value that would lead to exponential inflation. Finally, in \fig{theory4} the metastable branch extends all the way down to $T=0$. This indicates that this theory possesses two vacua, a stable one with zero energy and a metastable one with non-zero energy. Since both are Lorentz-invariant, the pressure at the metastable one is exactly $\mathcal{P}=-\mathcal{E} \neq 0$ and therefore $w=-1$.  The appearance of this metastable vacuum in the gauge theory is due to the appearance of an additional minimum in the scalar potential \eqn{potentialsuperpotential} on the gravity side 
\cite{Bea:2018whf}. 

The spinodal regions in \figs{theory2} and \ref{theory4} are slightly peculiar in that  the specific heat becomes positive in the central region (the red, dotted part). This may seem to indicate that this region is locally stable. However, analysis of the quasi-normal modes on the gravity side reveals that this region is actually dynamically unstable, 
along the lines of  \cite{Gursoy:2016ggq}. In any case, this peculiar feature could be removed by considering a more general potential with more parameters while still interpolating between models with and without thermal inflation. 

The two main messages from this section are as follows. First, supercooling does not necessarily lead to a period of thermal inflation. The former requires that bubble nucleation be sufficiently suppressed, whereas the latter requires that the metastable branch extend sufficiently close to $T=0$. Our model illustrates this distinction clearly. Since it corresponds to a large-$N$ gauge theory supercooling is always present, but whether thermal inflation  takes place depends on the choice of parameters. Second, irrespectively of whether thermal inflation does or does not take place, the phase transition can still be completed, since the Universe can eventually enter the spinodal branch, whose dynamics we describe in the following section.

\section{Initial state and time evolution}
\label{time}

We have argued above that, if the bubble nucleation rate is sufficiently suppressed, then the Universe will roll down the metastable branch and enter the spinodal region. Slightly past this point unstable modes can begin to grow. However, their growth rate, which is bounded from above by 
$\gamma_\textrm{max}$, is initially very small compared to the expansion rate of the Universe, since precisely at $T=\Ts$ we have $c_s=0$ and hence $\gamma_\textrm{max}=0$ --- see \fig{kandgamma}. Thus, the growth becomes important when the Universe has moved far enough along the spinodal branch so that $\gmax \sim H$. Let $\Tg$ be the ``growth temperature'' at which this happens. Since in the next few paragraphs we are only interested in parametric estimates we will not distinguish between the exact $\gmax, \kmax$, etc and their hydrodynamic approximations. Using the parametric dependence  \eqn{ball} we then have that $\Tg$ is determined by the condition 
\be
\label{when}
\gamma_\textrm{max} (\Tg) \sim  \left | c_s  (\Tg) \right |^2 \Tg \sim H \,.
\ee
In order to proceed we must distinguish three cases depending on the hierarchy between $T_g$ and $H$. We recall again that $\Tg$ refers to the temperature in the sector undergoing the phase transition, which may or may not coincide with the temperature in the sector driving the expansion, $\Trad$. 

If $H\gg T_g$ then the unstable modes do not have time to grow unless $\left | c_s \right |^2$ is parametrically large. This means that the Universe can traverse the spinodal region without the instability having a significant effect, so we will not consider this case further. 

If $\Tg \sim H$ then we must have $ \left | c_s  (\Tg) \right | \sim 1$ and the modes that will dominate the exponential growth will have momenta $\kmax$ that, by virtue of \eqn{ball}, is  given by 
\be
\label{sup3}
k_\textrm{max} \sim T_c \sim H \,,
\ee
where on the right-hand side we have approximated $\Tg \sim T_c$. A necessary condition for this case to occur is that the transition takes place in a hidden sector with $T_c \sim H\ll \Trad$. 

If the transition takes place in the sector driving the expansion then 
\mbox{$ T_g \sim \Trad$} and $H \ll \Tg$. In this case $\left | c_s  (\Tg) \right |^2 \ll 1$ and we can  use in \eqn{when} the approximation for the speed of sound  \eqn{csclose} to obtain 
\be
\label{Tg}
\Tg-\Ts \simeq \frac{\Hs^2}{\Ts}\,.
\ee
The time it takes for this change in temperature is parametrically small, since using \eqn{ht} we can translate \eqn{Tg} into 
\be
\label{tg}
\tg-\ts \simeq \frac{\Hs}{\Ts^2}\,.
\ee
Using again the parametric dependence \eqn{ball} we see that in this case the modes that will dominate the exponential growth will have momenta 
\be
\label{sup}
\kmax \simeq \sqrt{\frac{\Hs}{T_c}} \, T_c \,,
\ee
which satisfies
\be
\label{sup2}
H\ll \kmax \ll T_c \,.
\ee
where on the right-hand side of the last two equations we have approximated $\Tg \sim T_c$. 

In order to simulate the evolution numerically we need to choose a specific value of 
$\Tg$. As we will discuss below, we expect the results to be qualitatively similar for any choice. We will choose the value $\Tg=0.3908 \Lambda$, which differs from $\Ts$ by about 1\%. In other words, we choose the initial state to be the one indicated by the black circle in \fig{free}. For this state the hydrodynamic values of the key quantities introduced in \Sec{spinodal} are
\be
\label{paramhyd}
\kstarhyd=0.81\Lambda \sac \kmaxhyd=0.41\Lambda \sac 
\gmaxhyd = 0.073\Lambda \,.
\ee
We then imagine that small thermal fluctuations on top of this homogeneous state trigger the instability. Since the dynamics is dominated by the unstable modes, and since their typical momentum is $k \lesssim T$ according to \eqn{sup3} and \eqn{sup2}, we will ignore the Boltzmann factor $\exp(-k/T)$ and assume that fluctuations with any $k$ are equally likely in the initial state. For simplicity, in this paper we will assume that all initial fluctuations start off with the same small amplitude $\delta \mathcal{E}/\mathcal{E} = 10^{-4}$. We will come back to this point in  \Sec{disc}.

\begin{figure}[t!!!]
	\begin{center}
			\includegraphics[width=.8\textwidth]{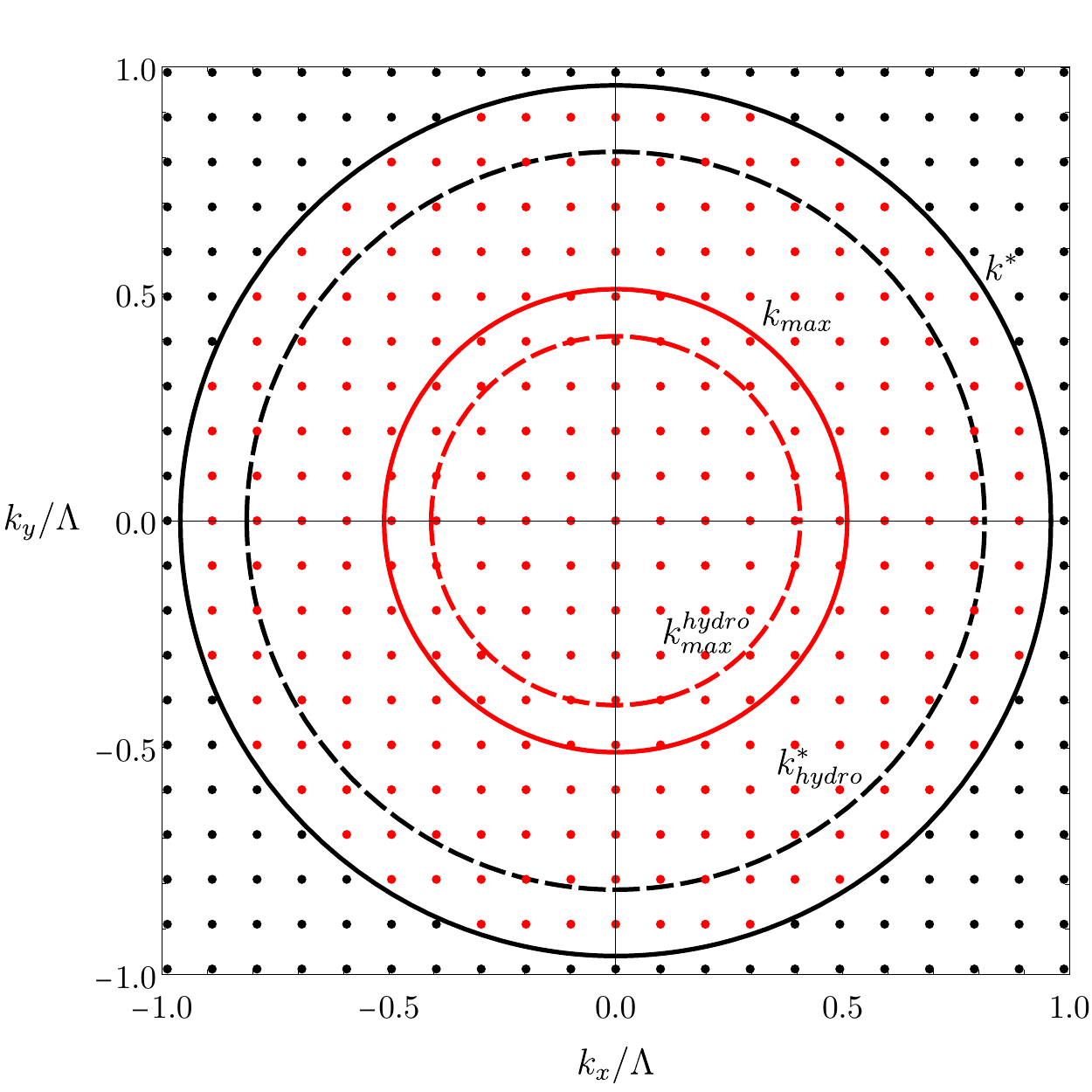}
		\caption{
		\label{kxky} 
		\small Allowed momenta in the $(k_x,k_y)$-plane. The solid curves indicate the exact values of $k_*$ (black) and $\kmax$ (red) for the initial state under consideration, given in \eqn{paramexact}.  The dashed curves indicate the hydrodynamic approximations to these quantities, $\kstarhyd$ and $\kmaxhyd=0$, given in \eqn{paramhyd}. Modes inside the black, solid circle are unstable. Within these, those with the largest growth rate are those on the solid, red circle.}
	\end{center}
\end{figure}

\begin{figure}[h!!!]
	\begin{center}
		\begin{tabular}{cc}
			\includegraphics[width=.45\textwidth]{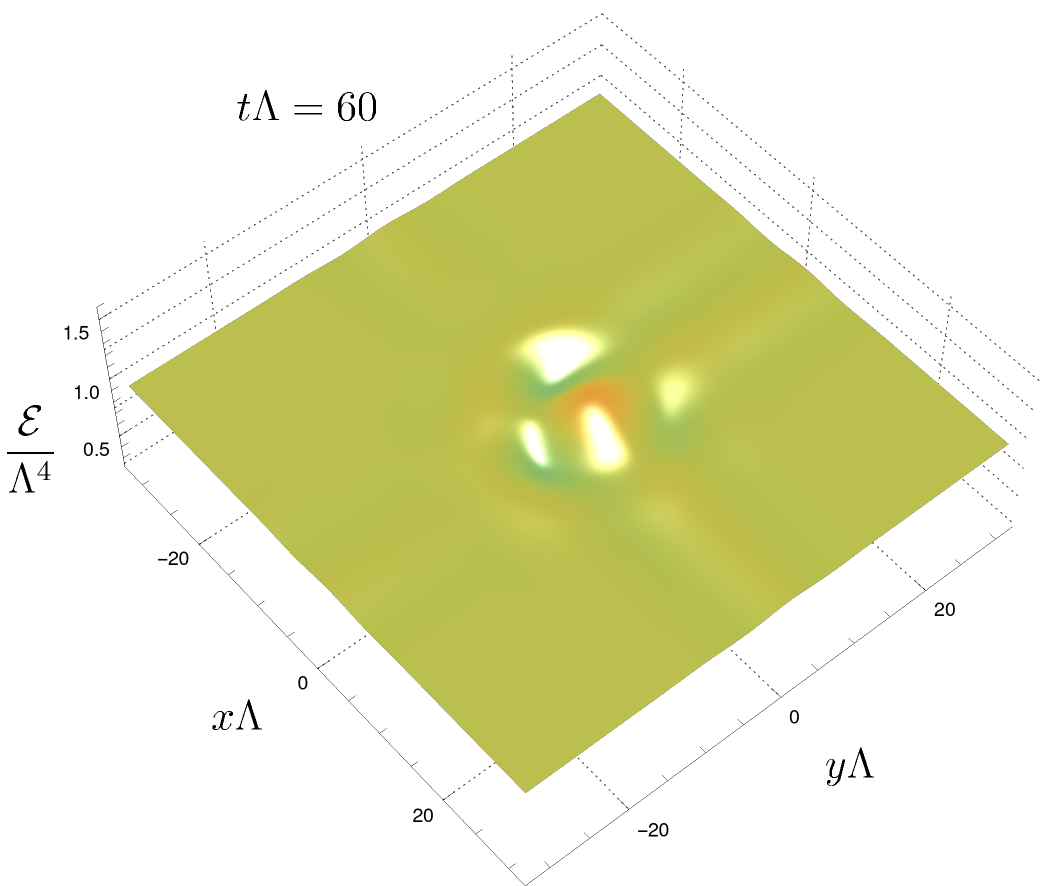}
			\,\,&
			\includegraphics[width=.45\textwidth]{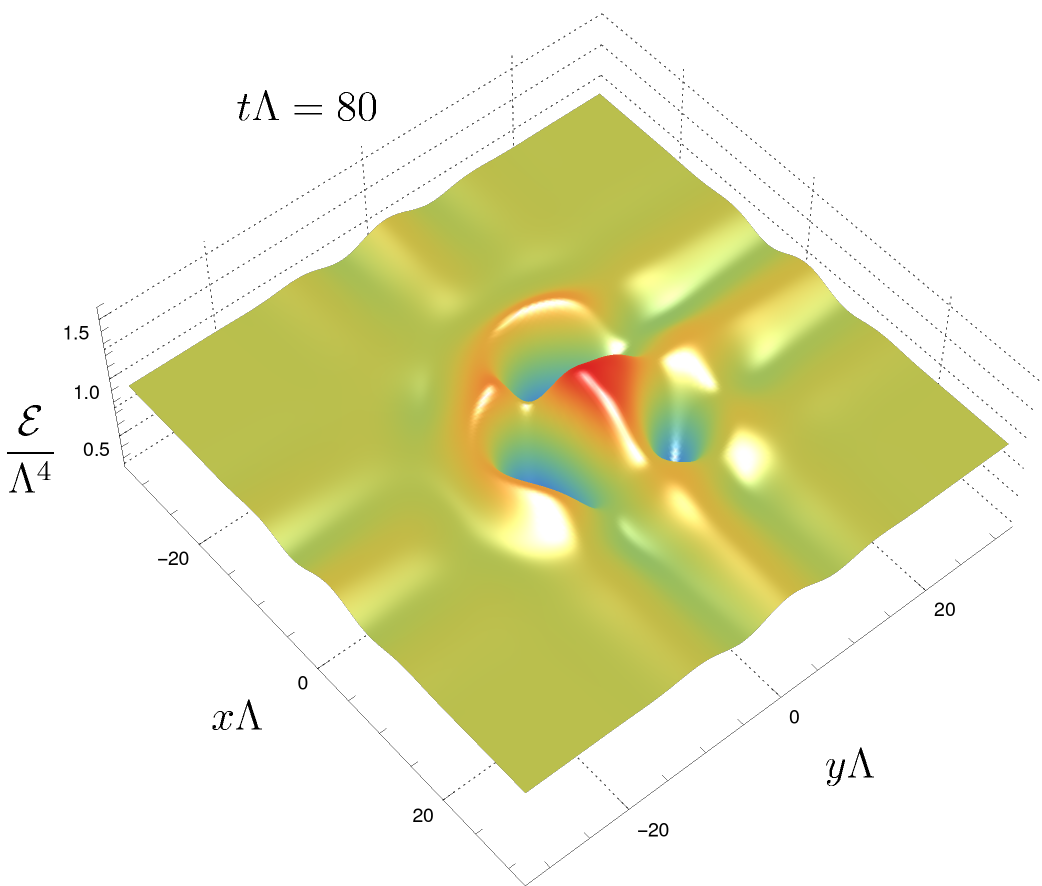}
			\\
			\includegraphics[width=.45\textwidth]{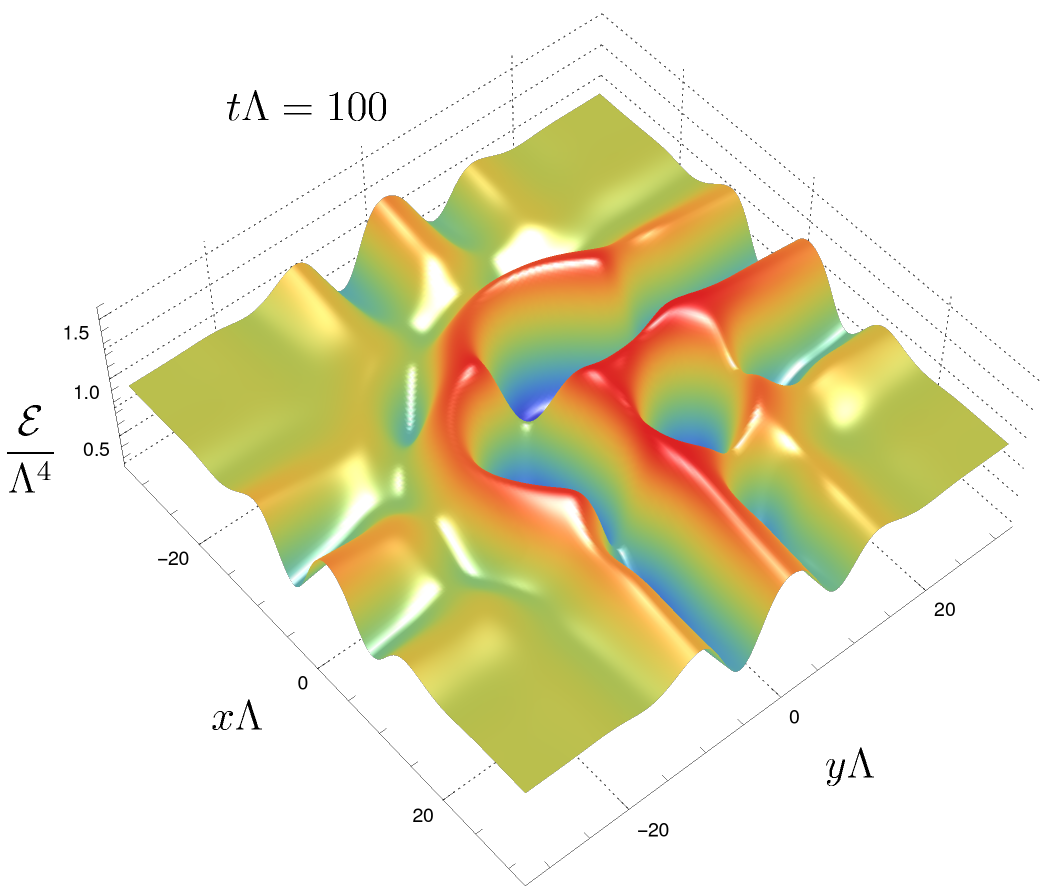}
			\,\,&
			\includegraphics[width=.45\textwidth]{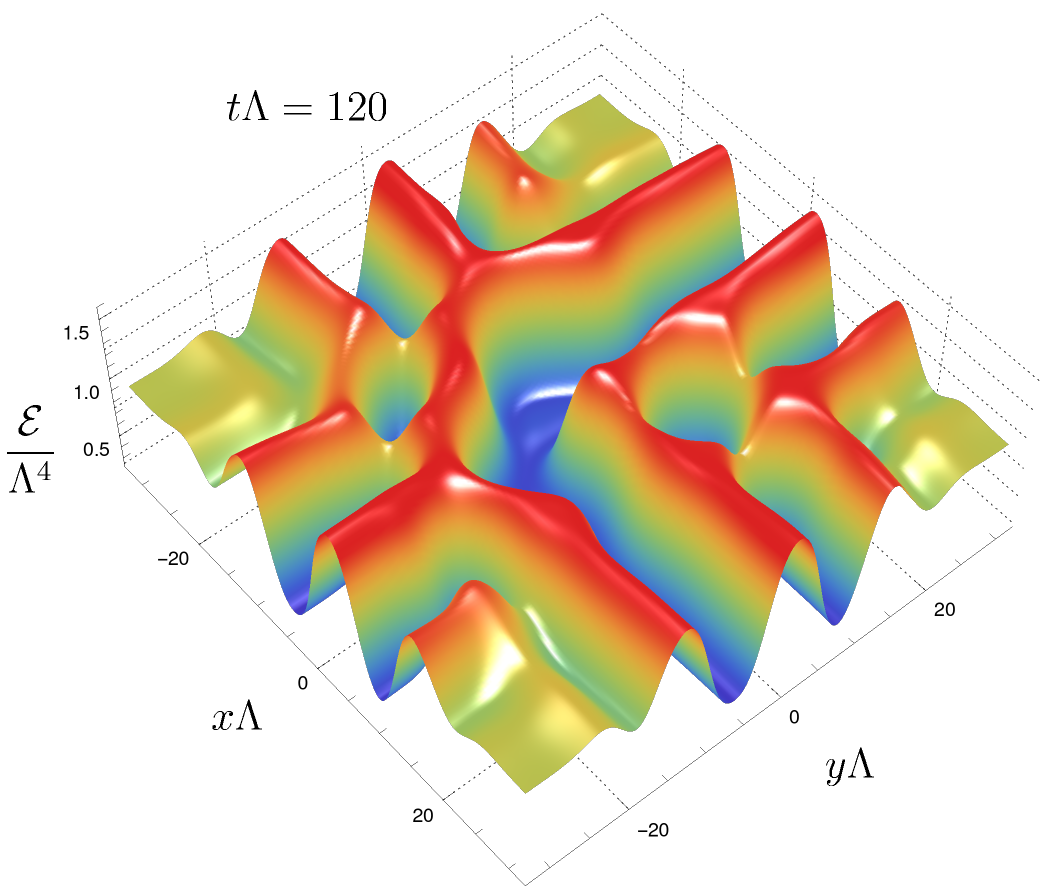}
			\\
			\includegraphics[width=.45\textwidth]{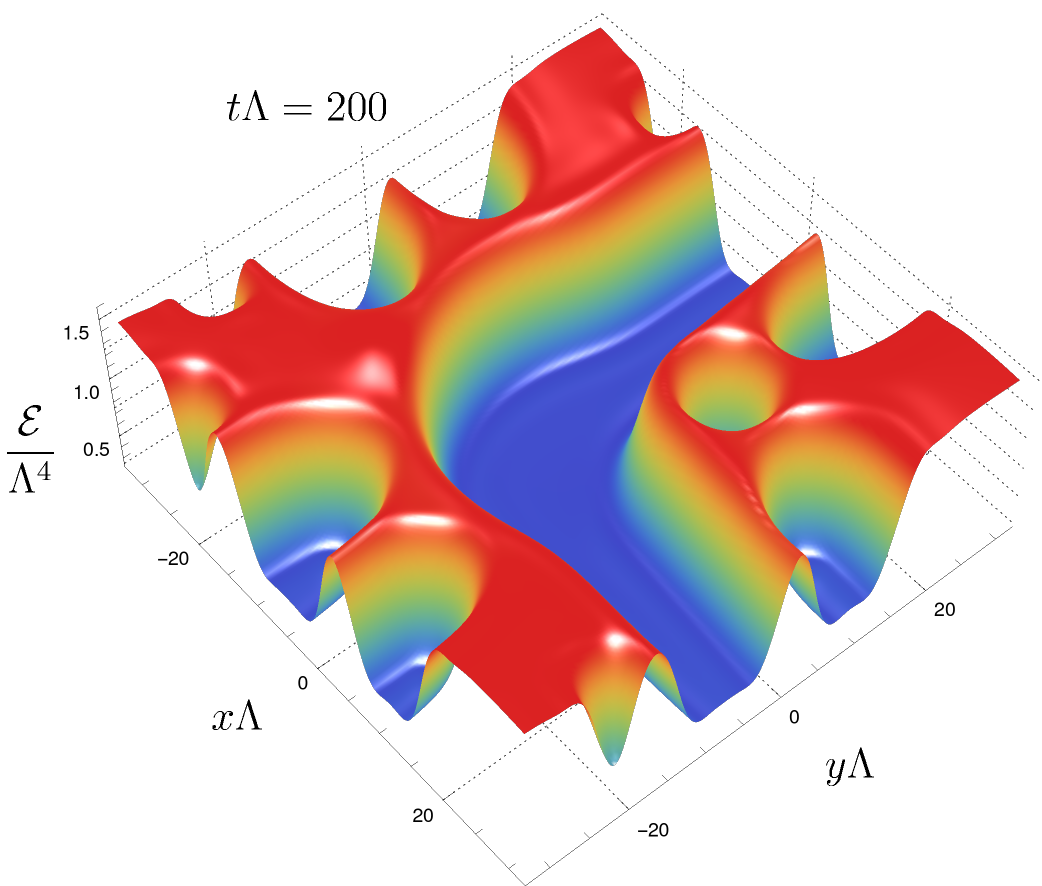}
			\,\,&
			\includegraphics[width=.45\textwidth]{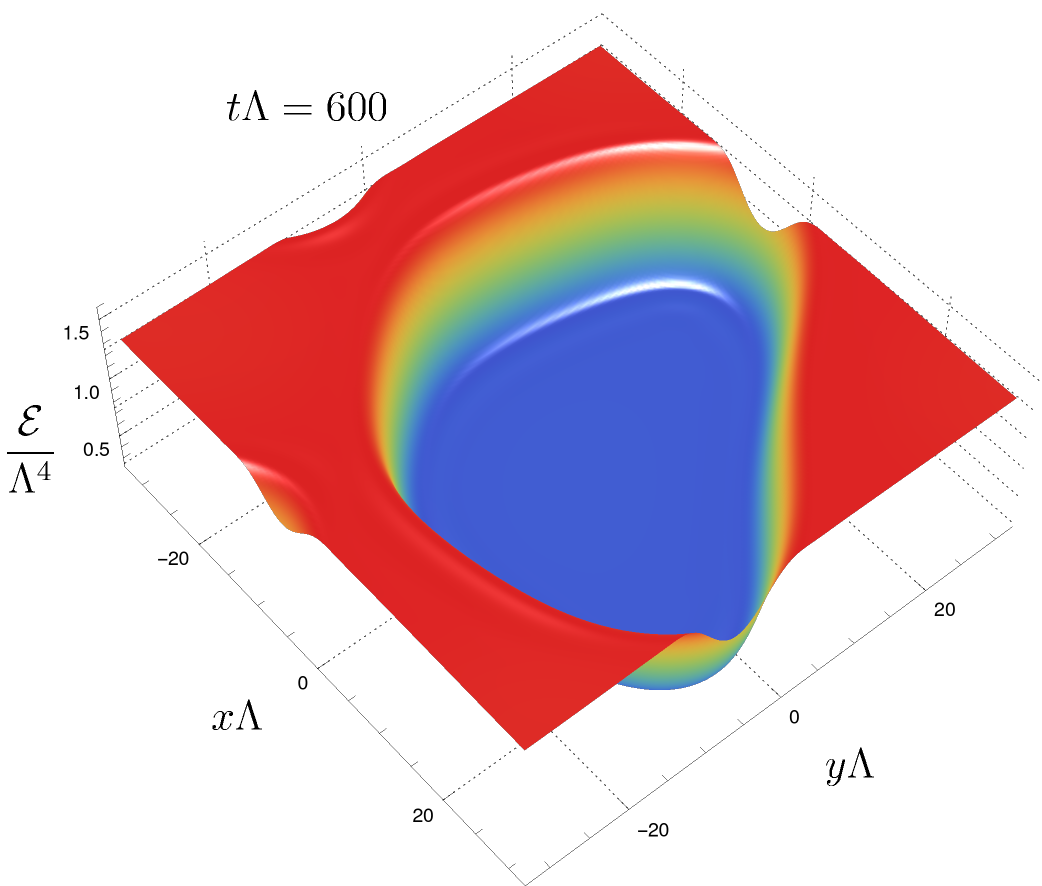}
		\end{tabular}
		\caption{\label{3Denergy} 	 \small Spacetime evolution of the energy density for the initial homogeneous state in the spinodal region indicated by a black circle in \fig{free}, perturbed with small fluctuations. A video of the evolution can be found at \href{https://www.youtube.com/watch?v=qIhbpchr3gE}{https://www.youtube.com/watch?v=qIhbpchr3gE}.
		}
	\end{center}
\end{figure}

As discussed in \sect{intro},  we will allow for dynamics only along two of the three spatial directions of the gauge theory, which we call $x,y$. In other words, we impose translational symmetry along $z$. In addition, we compactify the $x$- and the $y$-directions on circles of equal lengths 
\be
L=63.6 \Lambda^{-1} \,,
\ee
namely we impose periodic boundary conditions on these coordinates.  This infrared cut-off is technically convenient because it reduces the number of unstable sound modes to a finite number, since excitations along the $x$- and $y$-directions must have quantized momenta 
\be
\label{nn}
(k_x, k_y) = \frac{2\pi}{L} \, (n_x, n_y) \,,
\ee
with $n_x, n_y$ integer numbers.  This means that the allowed momenta lie on a two-dimensional grid in the 
$(k_x, k_y)$-plane, as shown in \fig{kxky}, where the  circles  indicate the values $k_*$ and $\kmax$ for the initial state under consideration. 
Although only modes with $n\leq 10$ are shown in the figure, we have included modes with up to $n=50$ in the fluctuations of the initial state, all of them with amplitude $\delta \mathcal{E}/\mathcal{E} = 10^{-4}$.  We have determined the corresponding time evolution by numerically evolving the Einstein-plus-scalar equations that follow from  \eqn{action}-\eqn{param} using our new code \cite{code}. From the near-boundary fall-off of the bulk fields we have extracted the boundary stress tensor. Snapshots of the resulting energy density  are shown in \fig{3Denergy}. The interested reader can also find a video of this evolution at \href{https://www.youtube.com/watch?v=qIhbpchr3gE}{https://www.youtube.com/watch?v=qIhbpchr3gE}.

\begin{figure}[t!!!]
	\begin{center}
			\includegraphics[width=.99\textwidth]{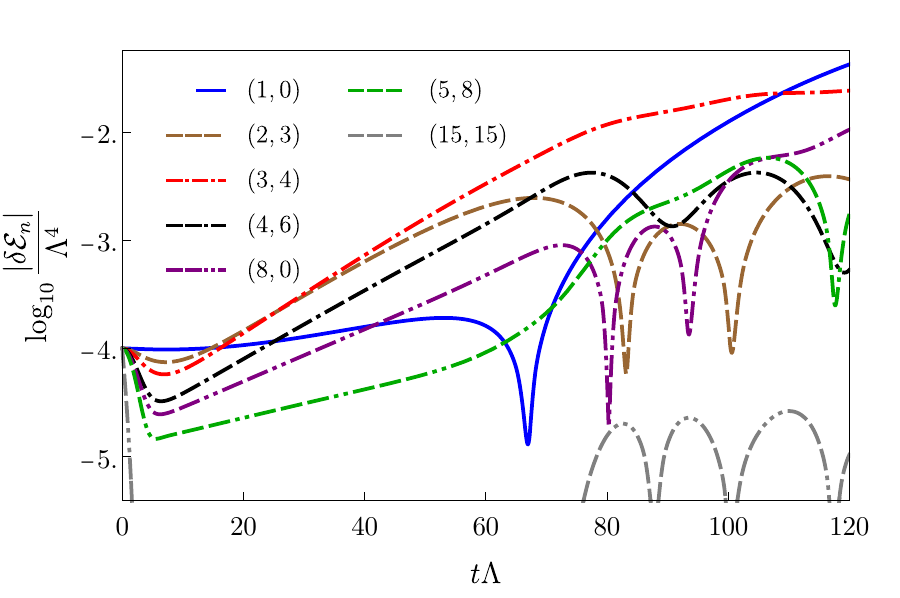}
		\caption{
		\label{modes} 
		\small Time evolution of the amplitudes of several Fourier modes corresponding to the run in \fig{3Denergy}. The different momenta are labelled by integers $(n_x, n_y)$ as in \eqn{nn}.}
	\end{center}
\end{figure}

As explained above, the initial state includes high-momentum fluctuations with momenta up to 
\be
|k| = 50 \times \frac{2\pi}{L} \simeq 5 k_* \,.
\ee
Most of these modes are outside the black, solid circle in \fig{kxky}, i.e.~they are stable. Since the initial perturbation is small, the first stage of the evolution is well described by a linear analysis around the initial homogeneous state. According to this, the stable modes 
decay exponentially fast as soon as the simulation begins, on a time scale of order $\Ts^{-1} \sim \Lambda^{-1}$.  For the unstable modes, linear theory predicts a behavior which is the sum of two exponentials, precisely the two solutions of the sound mode (\ref{disp}). In the spinodal region, one of these modes decays with time while the other one grows. After some time the latter dominates. This physics can be seen in \fig{modes}, which shows the time evolution of the amplitudes of several Fourier modes corresponding to the run in \fig{3Denergy}. The straight lines at early times correspond to the regime of exponential growth. At late times some of these slopes can change due to resonant behaviour, namely to the coupling between different modes \cite{Attems:2019yqn}.

\begin{figure}[t!!!]
	\begin{center}
			\includegraphics[width=.99\textwidth]{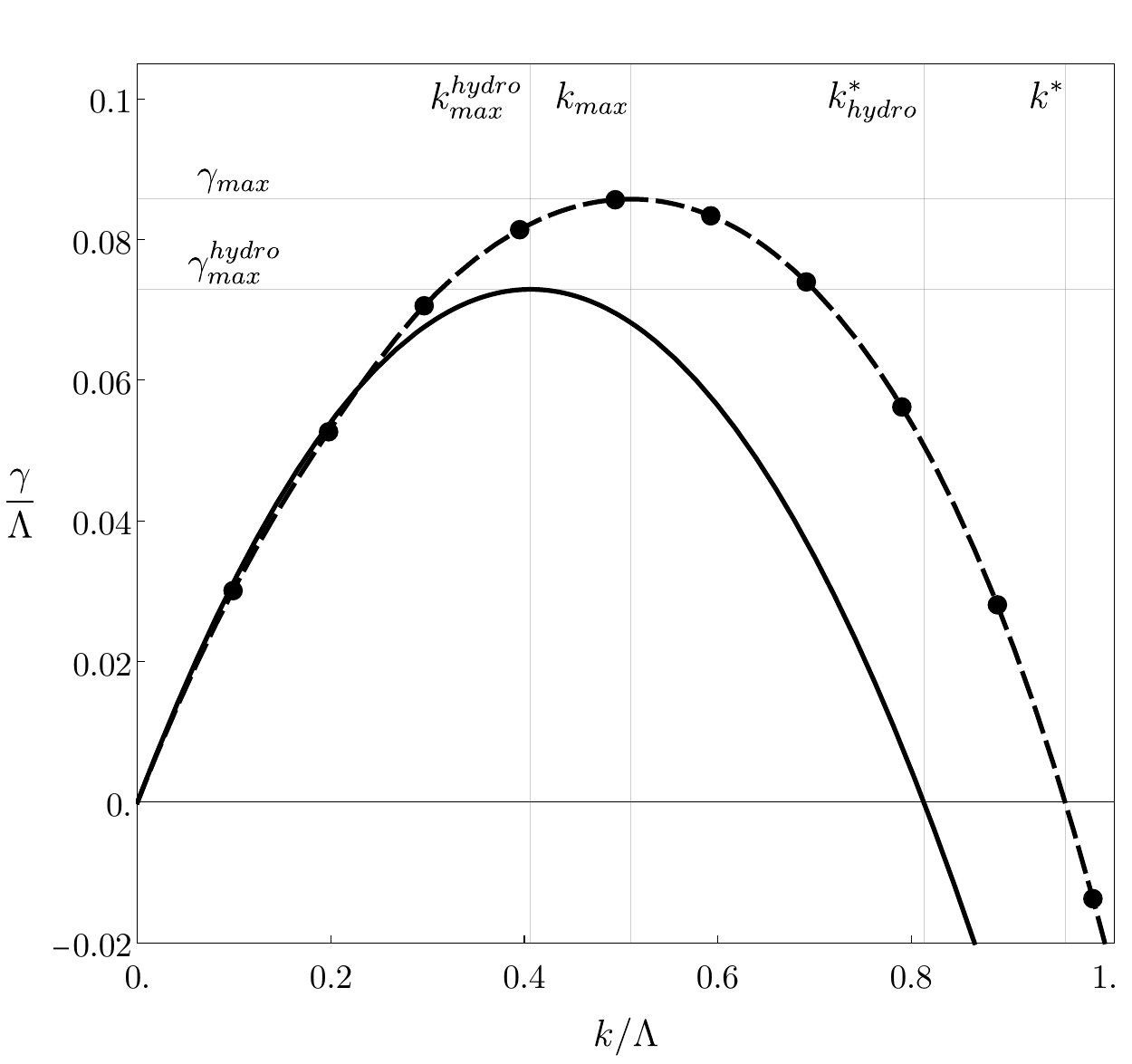}
		\caption{
		\label{gamma_true_k} 
		\small Comparison between dispersion relations. The continuous black curve shows the growth rates predicted by the hydrodynamic approximation (\ref{small}) for the state under consideration. The black dots show the growth rates extracted from a fit to the early-time slopes in \fig{modes}. The dashed black curve is an interpolation of these points.}
	\end{center}
\end{figure}

In \fig{gamma_true_k} we compare the growth rates predicted by the hydrodynamic approximation with those extracted from a fit to the slopes of the straight lines in \figs{modes} at early times. 
We see that the hydrodynamic approximation captures the correct qualitative shape everywhere, and that it provides a good approximation at the quantitative level at low $k$, as expected. The exact values of the key parameters, 
\be
\label{paramexact}
k_*=0.96\Lambda \sac \kmax=0.51\Lambda \sac 
\gmax = 0.086\Lambda \,,
\ee
are slightly larger than their  hydrodynamic counterparts.

Once the amplitudes of some  modes grow large enough, the evolution enters a non-linear regime. The dynamics here is rich and involves several phases. The interested reader can find a thorough discussion in \cite{Attems:2019yqn}. In a nutshell, the initial exponential growth gives rise to peaks and valleys. These are initially separated from each other by a distance of order $\kmax^{-1}$, but they eventually merge with one another until the system reaches a maximum-entropy state which, in large enough a box, is a phase-separated state at a constant, homogeneous temperature $T=T_c$. This dynamics can be observed by following the different snapshots in \fig{3Denergy}. In particular, the bottom-right plot shows a configuration that is close to a phase-separated state, in which the total volume is divided in two regions with energies precisely equal to $\Ehigh$ and $\Elow$, respectively, separated by an interface of thickness $\sim T_c^{-1} \sim \Lambda^{-1}$.

Note that in our simulations we have ignored the expansion of the Universe. We will come back to this point in \sect{disc}.

\section{Gravitational wave spectrum}
\label{gw}

We will consider GWs as metric perturbations around flat space, which ignores the expansion of the Universe. We will come back to this approximation in \sect{disc}. A GW  in flat space is described by a metric perturbation $h$ of the form
\be
ds^2=-dt^2 + (\delta_{ij} + h_{ij}) \, dx^i dx^j \,.
\ee
In this section, latin indices $i, j, m, n$ are spatial indices ranging from 1 to 3. Indices on $h_{ij}$ are raised and lowered with $\delta_{ij}$. In the  purely transverse traceless (TT) gauge, its evolution is governed by the perturbed Einstein equations
\be
\ddot{h}_{ij} (t, {\pmb{x}}) - \nabla^2 h_{ij} (t, {\pmb{x}}) = 16\pi G \, \Pi_{ij} (t, {\pmb{x}})\,,
\ee
where ${\pmb{x}}=(x^1,x^2,x^3)$ and  $\Pi_{ij}$ is the TT part of the stress tensor. In momentum space (along the spatial directions) the equation of motion becomes
\be
\label{eomk}
\ddot{h}_{ij} (t, \pmb{k}) + k^2 h_{ij} (t, \pmb{k}) = 16\pi G \, \Pi_{ij} (t,\pmb{k})\,,
\ee
where $\pmb{k}$ is the spatial three-momentum. 
The fact that $\Pi_{ij}$ is TT guarantees that it only sources the GW components of $h$, namely the spin-two, tensor component.   Given the stress tensor, its TT part is given by
\be 
\Pi_{ij} (\pmb{k}) = \Lambda^{mn}_{ij} (\hat{\pmb{k}}) \, T_{mn} (\pmb{k}) \,,
\ee
where a `` $\hat{}$ '' indicates a unit vector along the corresponding direction, 
\be
\Lambda^{mn}_{ij} (\hat{\pmb{k}}) = P^m_i (\hat{\pmb{k}}) P^n_j (\hat{\pmb{k}}) - 
\frac{1}{2} P_{ij} (\hat{\pmb{k}}) P^{mn} (\hat{\pmb{k}}) 
\ee
and
\be
\label{pp}
P_{ij} (\hat{\pmb{k}}) = \delta_{ij} - \hat{k}_i \hat{k}_j \,.
\ee

We have imposed translation invariance along the $z$-direction, so $\Pi_{ij} (t,\pmb{k})$ vanishes when 
$\pmb{k}$ points along this direction. Through \eqn{eomk} this means that we cannot produce GWs along this direction. Therefore, let us consider GWs propagating along a direction in the $xy$-plane at an angle $\theta$ with the $x$-axis. In this case 
\be
\label{kk}
{\pmb{k}} = (k_x, k_y , 0) 
\ee
and therefore the projector takes the form
\be
P_{ij} = \left(
\begin{array}{cc}
M  &  0  \\
 0 &  1  
\end{array}
\right) \sac
M = \frac{1}{k^2} \left(
\begin{array}{cc}
k_y^2  &  k_x k_y   \\
 k_x k_y &  k_x^2 
\end{array}
\right)\,.
\ee
In our case the stress tensor is 
\be
T_{ij}=\left(
\begin{array}{ccc}
T_{xx}  &  T_{xy}  &  0 \\
 T_{xy} &  T_{yy} & 0  \\
0  & 0  &   T_{zz}
\end{array}
\right) \,.
\ee
Projecting with $P$ we obtain
\be
\label{pipi}
\Pi_{ij} = A \times
\left(
\begin{array}{cc}
M  &  0  \\
 0 &  -1  
\end{array}
\right) \,,
\ee
where 
\be
\label{pi}
A=\frac{1}{2k^2}\Big( k_y^2 \, T_{xx} + 2 k_x k_y \, T_{xy} + k_x^2 \, T_{yy} - k^2 T_{zz} \Big)  \,.
\ee
The  tensor $\Pi_{ij}$ will source the  GW components 
$h_{xx}, h_{xy}, h_{yy}$ and $h_{zz}$. These will only propagate in the $xy$-plane, namely their dependence will be of the form $e^{i k _x x + i k_y y}$. 

Once $h$ has been found, the GW energy density is given by 
\be
\rhogw= \frac{1}{32\pi G L^3} \int d^3 x \, \dot{h}^{ij} (t,\pmb{x})\, \dot{h}_{ij} (t,\pmb{x}) \,,
\ee
where $L^3$ is the volume over which the energy is averaged. Integrating in $z$ and  Fourier-transforming in $x,y$ we get
\be
\rhogw=  \int  dk_x \, dk_y \, \frac{d \rhogw}{dk_x  dk_y} \,, 
\ee
where the energy density per unit two-momentum is
\be
\label{difTwo}
\frac{d \rhogw}{dk_x  dk_y} = \frac{1}{32\pi G L^2} \frac{1}{(2\pi)^2}\, 
\dot{h}^{ij} (t,k_x,k_y)\, \dot{h}^*_{ij} (t,k_x,k_y) \,.
\ee
Changing to polar coordinates $(k,\theta)$,  we define the differential energy density per unit logarithmic momentum as
\be
\label{dif}
\frac{d\rhogw}{d\log k} = k \frac{d\rhogw}{d k} =  
\frac{1}{32\pi G L^2} \frac{k^2}{\left( 2\pi \right)^2} \int_0^{2\pi} 
d\theta \, \dot{h}^{ij} (t,k,\theta) \, \dot{h}^*_{ij} (t,k,\theta) \,.
\ee

It is often convenient (see e.g.~\cite{Garcia-Bellido:2007fiu}) to make use of the fact that  the relation between $\Pi_{ij}$ and $T_{ij}$, as well as the equation of motion for $h_{ij}$, are both linear. This means that, instead of evolving \eqn{eomk} for the variable $h$, we can evolve the following equation of motion 
\be
\label{eomu}
\ddot{u}_{ij} (t, \pmb{k}) + k^2 u_{ij} (t, \pmb{k})= 16\pi G \, T_{ij} (t,\pmb{k}) 
\ee
for the auxiliary variable $u$, which is nothing but \eqn{eomk} but sourced by the full stress tensor instead of by its TT part. This is useful because projecting requires going to Fourier space, which is time-consuming. Then we can obtain the GW metric at any desired time by applying the projector to the solution,
\be 
h_{ij} (\pmb{k}) = \Lambda^{mn}_{ij} (\hat{\pmb{k}}) \, u_{mn} (\pmb{k}) \,,
\ee
and the differential energy density takes the form
\be
\label{difU}
\frac{d\rhogw}{d\log k} =  
\frac{1}{32\pi G L^2} \frac{k^2}{\left( 2\pi \right)^2} \int_0^{2\pi} 
d\theta \, \dot{u}^{ij} (t,k,\theta) \, \Lambda^{mn}_{ij} (\hat{\pmb{k}}) \, 
\dot{u}^*_{mn} (t,k,\theta)\,.
\ee


In the rest of Sec.~\ref{gw}, boldface vectors will denote either a three-dimensional vector with vanishing $z$-component, as in \eqn{kk}, or simply a two-component vector in the $xy$-plane, e.g.~${\pmb{k}} = (k_x, k_y)$. Which one applies will be clear from the context.

\subsection{Sound waves}
Since the modes that drive the spinodal instability are sound modes, we expect that the evolution at early times may be well described by hydrodynamics.  In the next section we will perform a quantitative analysis of the GW production in the hydrodynamic approximation. In order to develop some intuition, in this section we will analyse the production when only two isolated sound waves  collide. 

Consider a small perturbation around a system in thermal equilibrium. Assume that the perturbation can be described within the hydrodynamic approximation, namely that it is controlled by fluctuations $\delta T$ and $\delta v^i$ of the hydrodynamic variables. In the case of the velocity field we have $\delta v^i=v^i$ since the velocity is zero in static equilibrium. Then the fluctuation in the (spatial part) of the stress tensor takes the form
\be
\label{Thydro}
\delta T^{ij} = \delta^{ij}\,  c_s^2 c_v \, \delta T +  \omega_0 \, v^i v^j  \,,
\ee
where $\omega_0 = \mathcal{E}_0 + \mathcal{P}_0$ is the enthalpy of the unperturbed state, $c_v$ is the specific heat \eqn{heat},  and $c_s$ is the speed of sound \eqn{cs}. In writing \eqq{Thydro} we have worked only to second order in the fluctuations, in particular in velocities. This means that, in addition to the hydrodynamic expansion in gradients, we are further expanding in the amplitude of the fluctuations. We expect this to be justified at sufficiently early times. 

The first term in \eqn{Thydro} is a pure trace and therefore it does not contribute to the TT part of the stress tensor. This can be seen explicitly in \eqn{pipi}, since in this case 
\be
T_{xx}=T_{yy}=T_{zz} \sac T_{xy}=0 \,, 
\ee
and therefore the prefactor in this equation vanishes. Thus only the second term in \eqn{Thydro} can contribute to the production of GWs at leading order. The velocities can be extracted from the $T^{0i}$ components, since at leading order 
\be
\delta T^{0i} = \omega_0 v^i \,.
\ee

The fluctuations in the velocity field decompose into two decoupled channels. The longitudinal or sound channel is the one for which the momentum of the perturbation is aligned with the velocity field, namely $\pmb{k}$ is parallel to $\pmb{v}$. The transverse or shear channel is the one in which the  momentum of the perturbation is orthogonal to the velocity field. We will focus on the sound mode because this is the one that is unstable in the spinodal region. Before we proceed note that we need a superposition of at least two sound waves in different directions in order to produce GWs. Indeed, suppose we have a single sound wave with momentum $\pmb{q}$. Without loss of generality we can  assume that it propagates along the $x$-direction. Then
\be
\pmb{v}=(v^x,0,0) \sac \pmb{q}=(q_x,0,0) \,.
\ee
This means that the only non-zero component of the second term in \eqn{Thydro} is $\delta T^{xx}$. Moreover, the momentum of the resulting gravitational wave is $\pmb{k}=2\pmb{q}$, which also points along the $x$-direction. Therefore in \eqn{kk} we have $k_y=0$. Substituting this in \eqn{pi} we see that $A=0$. Physically, the reason for this is that a sound wave has spin zero and hence it only induces fluctuations in the pressure along its direction of propagation, whereas a GW is transverse, and therefore it is only sourced by fluctuations in the transverse components of the pressure. 

Consider therefore a superposition of two sound waves with momenta $\pmb{p}$ and $\pmb{q}$, and with velocities $\pmb{u}_p$ and  $\pmb{u}_q$ parallel to the respective momenta. Then the velocity field takes the form
\be
\pmb{v} (t,\pmb{x}) = \pmb{u}_p \, e^{\gamma_p t} \, e^{i \pmb{p} \cdot \pmb{x}} +
\pmb{u}_q \, e^{\gamma_q t} \, e^{i \pmb{q} \cdot \pmb{x}} \,,
\ee
where we have assumed that the time dependence is the one dictated by the spinodal instability with the corresponding growth rates $\gamma_p=\gamma(p)$ and 
$\gamma_q=\gamma(q)$. This velocity field  can source a GW with momentum along $\pmb{k}=\pmb{p}+\pmb{q}$. Without loss of generality, let us assume that $\pmb{p}$ points along the $x$-direction and that $\theta$ is the angle in the $xy$-plane between $\pmb{p}$ and $\pmb{q}$. Then
\bea
 \pmb{p} &=& (p,0,0) \,,\\ 
 \pmb{u}_p &=& (u_p , 0 ,0) \,,\\ 
 \pmb{q} &=& q (\cos\theta, \sin \theta,0) \,,\\  
 \pmb{u}_q &=& u_q (\cos\theta, \sin \theta,0) \,,\\
  \pmb{k} &=& (p + q \cos\theta, q \sin\theta ,0) \,.
\eea
The second term in \eqn{Thydro} has three contributions proportional to $\pmb{u}_p^2$, to $\pmb{u}_q^2$ and to the crossed term $\pmb{u}_p\pmb{u}_q$. The first two terms do not produce GWs for the same reason as with a single wave. Only the third one contributes. The corresponding non-zero components of the stress tensor become
\bea
\delta T_{xx} &=& \omega_0\, e^{\left( \gamma_p + \gamma_q \right) t } e^{i \pmb{k} \cdot \pmb{x}} \, u_p u_q \, \cos \theta \,, \\
\delta T_{xy} &=& \omega_0\, e^{\left( \gamma_p + \gamma_q \right) t } e^{i \pmb{k} \cdot \pmb{x}} 
\,  u_p u_q  \, \sin \theta\,. 
\eea
The prefactor in \eqn{pipi} is then 
\be
\label{pre}
A=\frac{1}{2 } \, \omega_0\, e^{\left( \gamma_p + \gamma_q \right) t } e^{i \pmb{k} \cdot \pmb{x}} \, u_p u_q  \, f(r,\theta) \,,
\ee
where $r=q/p$ and
\be
\label{f}
 f(r,\theta)=\frac{r \sin^2\theta(2+3 r \cos \theta)}{1+r^2+2r\cos\theta}  \,.
\ee

\begin{figure}[t!!]
	\begin{center}
		\begin{tabular}{cc}
			\includegraphics[width=.49\textwidth]{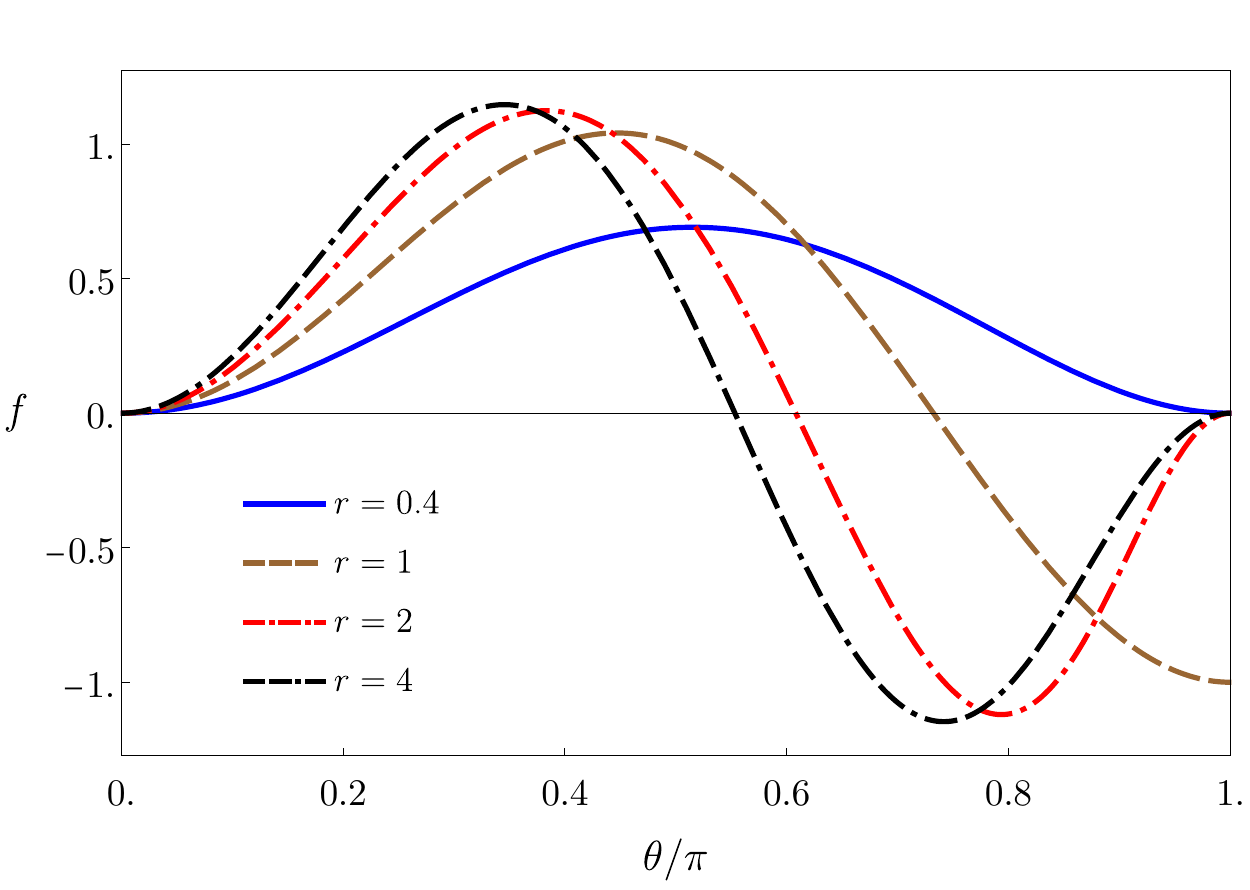}
			\,\,&
			\includegraphics[width=.49\textwidth]{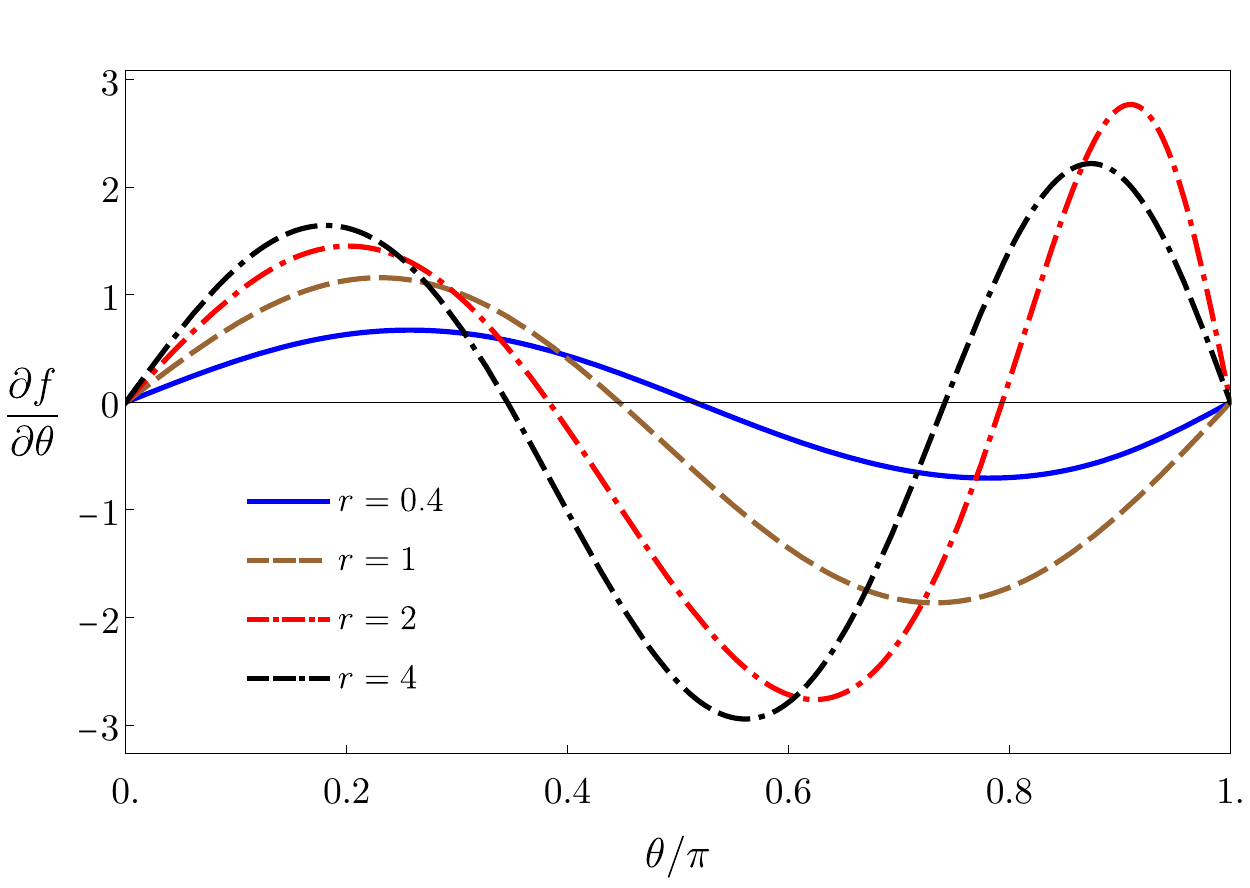}
		\end{tabular}
		\caption{\label{ffig} \small The function $f(r,\theta)$ (left) and its derivative $\partial f/\partial \theta$ (right).}  
	\end{center}
\end{figure}

The function $f(r,\theta)$ and its derivative $\partial f/\partial\theta$ are shown in Fig.~\ref{ffig}. We see that $f$ vanishes at $\theta=0$ regardless of the value of $r$. This is expected since in this case $\pmb{p}$ and $\pmb{q}$ are collinear. If $r\neq 1$ then $f$ also vanishes at $\theta=\pi$. However, if $r=1$ then $f\to -1$ as $\theta\to \pi$. This case corresponds to the limit $k\to0$, which is approached as  $\pmb{p}$ and $\pmb{q}$ become antiparallel.  

\eqq{pre} contains a crucial insight about the spectrum of GWs in the early-time part of the evolution. The key point is that the time dependence at a given GW momentum $\pmb{k}$ is controlled by the exponent $\gamma_p + \gamma_q$. The largest possible value of this exponent is $2\gmax$, with $\gmax$ the maximum growth rate in \fig{gamma_true_k}. This maximum value can only be realised by composing two sound-wave momenta  $\pmb{p}$ and $\pmb{q}$ such that their moduli satisfy $p=q=\kmax$. In other words, $\pmb{p}$ and $\pmb{q}$ must lie on the red, solid circle in \fig{kxky}. By adding two such momenta one can only obtain momenta $\pmb{k}$ with modulus in the range 
\be
\label{range}
0\leq k\leq 2\kmax \,. 
\ee
As a result, the GW spectrum  \eqn{dif} will be exponentially suppressed for $k\geq 2\kmax$. The fact that $f$ does not vanish in the opposite limit, $\k\to 0$, means that the TT part of the stress tensor is not suppressed in this limit. Therefore we expect that the only suppression of the differential energy density $d\rhogw/d\log k$ in the limit $k\to 0$ will be due to the $k^2$ factor in \eqn{dif}.

\subsection{Hydrodynamics}
We will now provide a more detailed analysis of the GW production in the early-time part of the evolution based on hydrodynamics. In this approximation, the part of the perturbed stress tensor that gives rise to the TT component is the second term in \eqn{Thydro}, namely
\be
\delta T_{ij}(t, \pmb{x})= \omega_0 \,  v_i (t, \x) v_j (t, \x) \,. 
\ee 
In Fourier space, this leads to 
\be
\delta T_{ij}(t, \k)=  \we_0 \int \frac{d^2 q }{(2\pi)^2} \, v_i (t, \k-\q) \, v_j(t, \q) \,.
\ee
Note that, in this and in some subsequent equations, boldface symbols such as $\x, \k, \ldots$ refer to two-dimensional vectors. 

Since the excitations of interest are sound waves, the velocity is longitudinal, which means that 
\be
\v (t, \q)= v(t, \q) \, \hat \q \,.
\ee
Using the hydrodynamic equations of motion, we can relate the modulus of the velocity field to the time derivative of the energy density fluctuation:
\be
v(t, \q)= \frac{i}{q \, \omega_0}\,  \partial_t  \, \delta \E (t,\q)  \,.
\ee
In our simulation the initial velocity at $t=0$ vanishes, and the amplitude of the energy density perturbation is a small number $\epsilon_{\q}$. This initial value arises as a combination of the growing and the decaying sound modes. Imposing the condition 
$\v (0, \q)=0$ fixes the relative coefficient in such a way that 
\be
\label{decaying}
\delta \E (t,q) =\frac{\epsilon_{\q} }{2}\left[
 \left(1 +\frac{\hq}{2} \right) e^{\gamma_q t}  +
 \left(1 -\frac{\hq}{2} \right) e^{-\tilde \gamma_q t} 
 \right] \,,
\ee
where $\gamma_q$ and $\tilde \gamma_q$ are the growth rates of the unstable and stable modes, respectively. 
In addition, for any vector of modulus $q$ we define 
\be
\hq\equiv\frac{q}{\kmax} \,.
\ee
Note that both $\delta \E (t,q)$ and $\epsilon_{\q}$ have dimensions of (mass)$^2$ because of the Fourier transform. In hydrodynamics, the growth rates are given by
\be 
\gamma_q =  \frac{|c_s|}{2} \, \hq (2-\hq) \, \kmax \sac 
\tilde \gamma_q =  \frac{|c_s|}{2} \, \hq (2+\hq) \, \kmax \,,
\ee
as can be seen by using the first equation in \eqn{why} to eliminate $\kmax$ from these equations. 

Hereafter we will only consider the unstable mode, since it dominates as soon as the stable mode has decayed. With this approximation we can write
\be
\label{here}
\delta T_{ij}(t, \k)= -\frac{|c_s|^2}{4 \we_0} \int \frac{ q \, d q \, d\theta  }{(2\pi)^2}  \, \epsilon_{\q} \epsilon_{\q'} \,  \left( 1-\frac{\hq^2}{4}\right)    \left( 1-\frac{\hq^{'2}}{4}\right) \frac{q_i }{q}\frac{q_j'}{q'} \, e^{(\gamma_q + \gamma_{q'}) t} \,,
\ee
where $\theta$ is the angle between $\k$ and $\q$ and 
\be
\q'=\k-\q \sac q' = \sqrt{k^2 + q^2 -2 k q \cc} 
\,.
\ee
Without loss of generality we assume that $\k$ points along the $x$-direction. 
The exponent controlling the time growth in the integrand of \eqn{here} takes the form
\be
t \gtot (k,q,\theta)= t \left( \gamma_q + \gamma_{q'} \right)= 
t \, \kmax \, \frac{|c_s|}{2} \Big[  \hq (2-\hq) + \hq' (2-\hq') \Big] \,.
\ee
If 
\be
\label{exist}
k \leq 2\kmax 
\ee
then, for fixed $k$, this exponent has a maximum at 
\be
\label{virtue1}
 \tilde q_M= 1 \sac \cc_M= \frac{\hk}{2 } \,,
\ee
and at this point we have 
\be
\label{virtue2}
\tilde q_M'= 1 \sac \gamma_{\textrm{tot}, M} = 2\gmax \,.
\ee
These equations mean that, for momenta $\k$ in the range \eqn{range}, the exponent achieves its largest possible value $2t\gmax$ and the integral is dominated by sound waves with momenta of equal moduli, 
\be
\label{equal}
q=q'=\kmax \,, 
\ee
lying at angles $\theta_M$ and $-\theta_M$ with respect to $\k$. This reproduces the conclusion anticipated in the paragraph of  \eqn{range}. In this range we can evaluate the integral via a saddle-point approximation. For this purpose, we expand the exponent to quadratic order around its maximum. The Hessian in variables 
\be
\Delta q= q-q_M \sac 
\Delta \cc=\cc-\cc_M 
\ee
 is not diagonal. Therefore, to facilitate the Gaussian integration we introduce the variables $(\alpha, \beta)$ defined through 
\be
\Delta q= \kmax \left( \alpha \cos\Psi   +   \beta \sin\Psi  \right)  \sac 
\Delta \cc= -\alpha \sin\Psi  + \beta \cos \Psi  \,,
\ee
with 
\be
\label{psi}
\tan 2\Psi=  \frac{4 \hk (2-\hk^2)}{(8-8\hk^2+\hk^4)} \,. 
\ee
With these variables we can write the exponent up to quadratic order as
\be
t \gtot(k,\alpha,\beta) \, \simeq \, t |c_s| \kmax  - \frac{\alpha^2}{2 \sigma^2_\alpha} - \frac{\beta^2}{2 \sigma^2_\beta} \,,
\ee
with 
\bea
\sigma^2_\alpha&=&\frac{1}{t |c_s|  \kmax  \left( 1 + \frac{\hk^4}{8} + \sqrt{1-\hk^2 + \frac{\hk^4}{4} +  \frac{\hk^8}{64}}
\right) } \,, \\[2mm]
\sigma^2_\beta &=& \frac{1}{t |c_s|  \kmax  \left( 1 +\frac{\hk^4}{8} - \sqrt{1-\hk^2 + \frac{\hk^4}{4} +  \frac{\hk^8}{64}} \right) } \,.
\eea
For generic values of $\hk$ these widths become arbitrary narrow as time increases 
and therefore we can replace their exponentials by $\delta$-functions  with appropriate
normalizations.  However, for small $\hk$ the  width of the $\beta$-variable scales as
\be
\sigma^2_\beta  \sim \frac{1}{t |c_s|  \kmax  \,\hk^2 } \,.
\ee
Therefore, the $\delta$-function approximation is only valid for 
\be
\label{rangerange}
\klow \ll k <2\kmax \,,
\ee
where 
\be
\label{klow}
\klow \equiv \sqrt{\frac{\kmax}{|c_s| t}}
\ee
and the upper bound comes from the condition \eqn{exist} for the existence of the saddle. Under these  conditions, we can write the stress tensor as 
\be
\label{focus}
\delta T_{ij}=- \frac{9}{64 \pi} \, \epsilon_{\q_M}  \epsilon_{\q'_M} \, \, 
\frac{|c_s|}{\we_0 \,   t}\,
\frac{\kmax}{\hk \sqrt{4-\hk^2}}  \,\,   \tilde q_{M\, i} \, \tilde q_{M\, j}' \, e^{2\gmax t}
\qquad \Big[ \klow \ll k \leq 2\kmax \Big]\,,
\ee
where 
\be
\tilde q_{M\, i}=  \left(\frac{\hk}{2}, \sqrt{1-\frac{\hk^2}{4}}, 0\right)\sac
\tilde q_{M\, j}'=  \left(\frac{\hk}{2}, -\sqrt{1-\frac{\hk^2}{4}}, 0\right) 
\,,
\ee
and we recall that we have assumed that $\k$ points along the $x$-direction.

For sufficiently small $k$ we see from \eqn{psi} that $\Psi \to 0$ and hence 
$\Delta q \simeq \kmax \alpha$ and $\Delta x \simeq \beta$. In this limit we can still perform the Gaussian approximation for the integral over $k$. This localises the integrand at $q=q'=\kmax$. Assuming that $\epsilon_{\q},  \epsilon_{\q'}$ are 
$\theta$-independent, the only $\theta$ dependence in the integrand is in the vectors
\be
\pmb{q}/q=(\cos \theta, \sin\theta, 0 ) \sac \pmb{q}'/q=(-\cos \theta, - \sin\theta, 0 )\,.
\ee
Integrating over $\theta$ we then find that, at leading order in $k$ in the limit $k\to 0$, the stress tensor approaches a finite, $k$-independent value
\be
\delta T_{ij}=- \frac{9}{256 \pi^{3/2}} \, \epsilon_{\q_M}  \epsilon_{\q'_M} \, \, 
\frac{|c_s|^{3/2}}{\we_0}\,
\frac{\kmax^{3/2}}{\sqrt{t}}  \,\,   q_{M\, i}' \, q_{M\, j} \, e^{2\gmax t}
\qquad \Big[ k \to 0 \Big] \,,
\ee
where 
\be
\tilde q_{M\, i} \, \tilde q_{M\, j}' =  -\mbox{diag}(\pi, \pi,0) \,.
\ee
We emphasize again that, as the vector $\k$ approaches zero, it does so along the $x$-direction. 

For $k> 2\kmax$ the exponent in \eqn{here} must be smaller than $2 \gmax$ since $k$ cannot be obtained by adding two vectors of modulus $\kmax$. The fact that the growth rate 
$\gamma(k)$ is a concave function of $k$, meaning that $\gamma''(k)<0$, implies that the maximum value of $\gtot$ is obtained by taking $\q$ and $\q'$ to be almost parallel to each other and to $k$ and to have equal moduli $q=q'=k/2$. They cannot be exactly parallel because then the TT component of the resulting stress tensor would vanish identically, but they can be exponentially close to being parallel. In this case the growth rate is given by 
\be
\label{expdecay}
\gtot \simeq 2 \, \gamma \left( \frac{k}{2} \right)  < 2\gmax \,.
\ee
Therefore, for $k> 2\kmax$, we expect the GW production to be exponentially suppressed with respect to the case $k< 2\kmax$ by a factor 
\be
\frac{\rhogw (k>2\kmax)}{\rhogw (k<2\kmax)} \simeq 
\exp \Big\{  - 4 t \Big( \gmax -  \, \gamma \left( k/2 \right)\Big) \Big\} \,.
\ee

Using the stress tensors above we can compute the spectrum of GWs. For concreteness let us focus on the case \eqn{focus}. Substituting this in \eqn{eomu} we find that, neglecting subleading terms in $1/t$, the solution is
\be
u_{ij} (t,\k)=-  \frac{9}{4}  \, \epsilon_{\q_M}  \epsilon_{\q'_M} \, \, 
 \frac{1}{\hk \sqrt{4-\hk^2}} \,\, \frac{1}{|c_s^2| + \hk^2}  \,\, \frac{ G |c_s|}{\we \, \kmax^3 t} 
 \,\,   \tilde q_{M\, i} \, \tilde q_{M\, j}' \, e^{2\gmax t} \,.
\ee
We now need to compute the contraction associated to the projector $\Lambda^{mn}_{ij}$ in \eqn{difU}. If we assume that $\epsilon_{\q_M},  \epsilon_{\q'_M}$ are independent of the angle  then $u_{ij} (t,\k)$ only depends on $k$ and we can assume that $\k$ points along the $x$-direction. In this case the projector \eqn{pp} becomes 
\be
P_{ij}=\mbox{diag} \left( 0, 1, 1 \right) \,.
\ee
Since $u_{ij}^* (t,\k) = u_{ij} (t,-\k)$ we  obtain
\be
\tilde q_{M}^i (\k) \, \tilde q_{M}^{' j} (\k) \, \Lambda^{mn}_{ij}(\hat \k) \, 
 \tilde q_{M\, m} (-\k)\, \tilde q_{M\, n}'(-\k) = \frac{1}{4}\sqrt{4-\tilde k^2}\,.
\ee
Substituting in \eqn{difU} we finally arrive at
\be
\label{valid}
\frac{d\rhogw}{d\log k} = \frac{G}{32\pi^2 L^2}\,
\left(\frac{9}{16}\right)^2 \,  
\epsilon_{\q_M} \epsilon_{\q_M'}  \epsilon_{-\q_M} \epsilon_{-\q_M'}\,  
\frac{1-\frac{\hk^2}{4}}{\Big( |c_s|^2+\hk^2 \Big)^2} \,\, 
\frac{ |c_s|^4 \kmax^2}{\omega_0^2 \, t^2}  \,\,  e^{4\gmax t} \,,
\ee
where we have made use of the fact that 
\be
\q_M (-\k) =\q'_M (\k)\,.
\ee
Recall that \eqn{valid} is only valid in the range  \eqn{rangerange}. To obtain an estimate for the total energy we proceed as follows. First, we note that, in a statistical ensemble characterised by a white noise, we expect that 
\be
\left<  \epsilon_{\q_M} \epsilon_{\q_M'}  \epsilon_{-\q_M} \epsilon_{-\q_M'} \right> = 
\left< \epsilon_{\q_M}   \epsilon_{-\q_M}   \right>^2 = \kappa \left( \kmax \right)^2 \,,
\ee
where $\kappa (q)$ is the strength of the noise and we have made use of \eqn{equal}. Second, the relevant integral is given by
\be
\int d \hk\,  \frac{1}{\hk}  \, 
\frac{1-\frac{\hk^2}{4}}{\left( |c_s|^2+\hk^2 \right)^2} 
 =\frac{1}{8 |c_s|^4}\left( \frac{|c_s|^2 (4+|c_s|^2)}{|c_s|^2+\hk^2} +
 4\log \frac{\hk^2}{c_s|^2+\hk^2}\right) \,.
\ee
Assuming that  
\be
|c_s|^2\ll \sqrt{ \frac{1}{ |c_s| \kmax t}} \ll 2 
\ee
we can expand the integral for small $|c_s| $. In this case the integral is dominated by the lower cut-off, and to leading order in the cut-off we get
\be
\int_{\sqrt{ 1 /(|c_s| \kmax t)}}^4 \, d \hk\,  \frac{1}{\hk}  \, 
\frac{1-\frac{\hk^2}{4}}{\left( |c_s|^2+\hk^2 \right)^2}   \simeq 
\frac{1}{4} \Big( |c_s| \kmax t \Big)^2 \,.
\ee
The energy density is thus given by
\be
\rhogw=\frac{G}{32\pi^2 L^2}\, \left(\frac{9}{32}\right)^2 \,  
\epsilon_{\q}^4 \,
 \frac{ |c_s|^6 \kmax^4}{\omega_0^2 \, }    \,\,  e^{4\gmax t} \,.
\ee
The end of the exponential regime is  determined by the condition that the amplitudes of the growing modes become of the same order as the latent heat \eqn{lat} \cite{Attems:2019yqn}. Thus
\be
\epsilon_{\q} \,   e^{\gmax t} \sim \frac{\Elat}{\kmax^2}  \,.
\ee
Substituting into the expression for the energy we arrive at
\be
\label{asin}
\rhogw=\frac{G}{32\pi^2 L^2}\, \left(\frac{9}{32}\right)^2\, \,
\frac{\Elat^4 \, |c_s|^6}{\omega_0^2\,  \kmax^4} \,.
\ee
The parametric dependence of this equation can be understood as follows. Conservation of the stress tensor implies that, at the end of exponential evolution, the velocity field is 
\be
v(x)\sim |c_s| \frac{ \Elat}{\omega_0} \,,
\ee
where we have used the fact that $\partial_t \sim |c_s| \kmax$. 
The relevant part of the stress tensor is then 
\be
T_{ij} (x) \sim \omega_0 \, v(x)^2 \sim  \frac{\Elat^2}{\omega_0}\,  |c_s| ^2 \,.
\ee
It follows that the metric fluctuation scales as
\be
 h_{ij}  (x )\sim  \frac{G}{\kmax^2} \, \frac{\Elat^2}{\omega_0}\,  |c_s| ^2 \,,
\ee
where we have used that the characteristic size of $T_{ij}$ is $1/\kmax$. Therefore 

\be
\dot h(x)_{ij}^2 \sim  \frac{G^2}{\kmax^2} \, \frac{\Elat^4}{\omega_0^2}\,  |c_s|^6 \,.
\ee
Since the characteristic size of  $\dot h$ is also $1/\kmax$, integrating over space and dividing by $G L^2$ we get 
\be
\rhogw \sim \frac{G}{L^2 \, \kmax^4} \, \frac{\Elat^4}{\omega_0^2}\,  |c_s|^6 \,,
\ee
as in \eqn{asin}.

\subsection{Full result}
In the previous section we estimated the GW emission in the regime of exponential growth. Once some modes become large enough this regime ends and the non-linear evolution begins \cite{Attems:2019yqn}. In this section we will present the exact results for the GW production for both the linear and the non-linear regimes. 

\begin{figure}[t!!]
\begin{center}
\includegraphics[width=.99\textwidth]{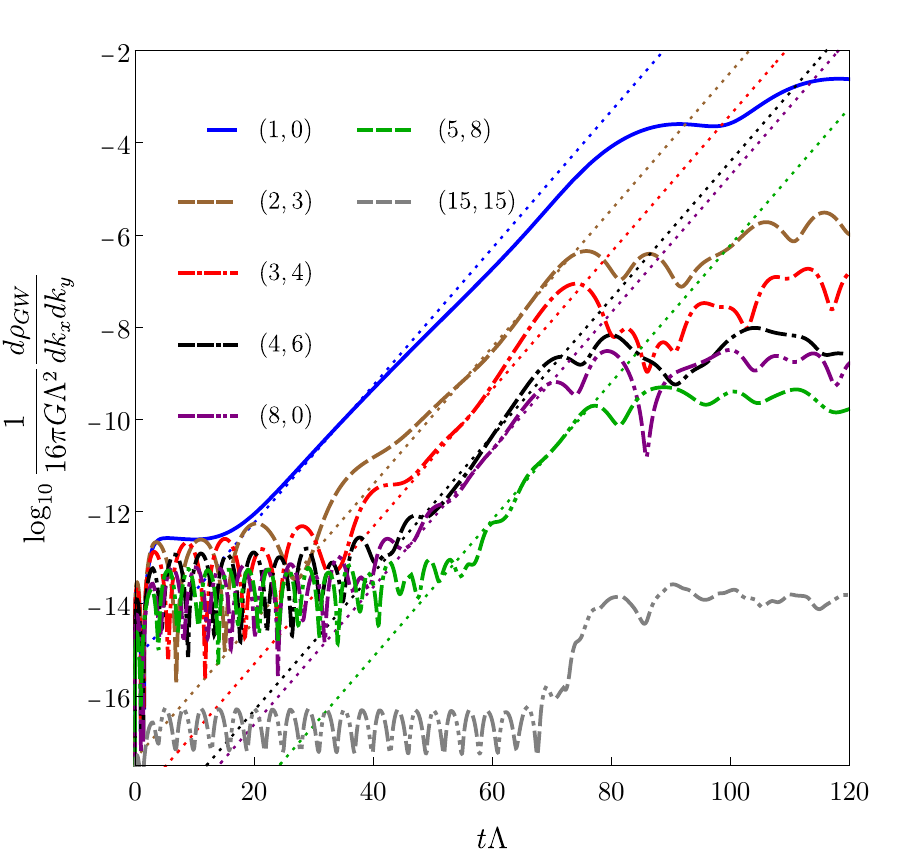}
			\caption{\label{fac} \small GW energy density per unit two-momentum \eqn{difTwo} at early times. The different momenta are labelled by integers $(n_x, n_y)$ as in \eqn{nn}. Except for the mode with $n_x=n_y=15$, all the other modes shown have $k<2\kmax$ and therefore at early times they grow approximately as $e^{4\gmax t}$, as indicated by the dotted, diagonal  lines. }
	\end{center}
\end{figure}

The behaviour of the differential energy density per unit two-momentum at early times is shown in \fig{fac}. 
We see that the energy density in modes with $k<2\kmax$ grows at early times approximately as $e^{4\gmax t}$, as predicted by \eqn{valid}. The deviations from this  behaviour, which begin around $60\lesssim t\Lambda \lesssim 80$, indicate the end of the linear regime.

\begin{figure}[t!!]
\begin{center}
\includegraphics[width=.99\textwidth]{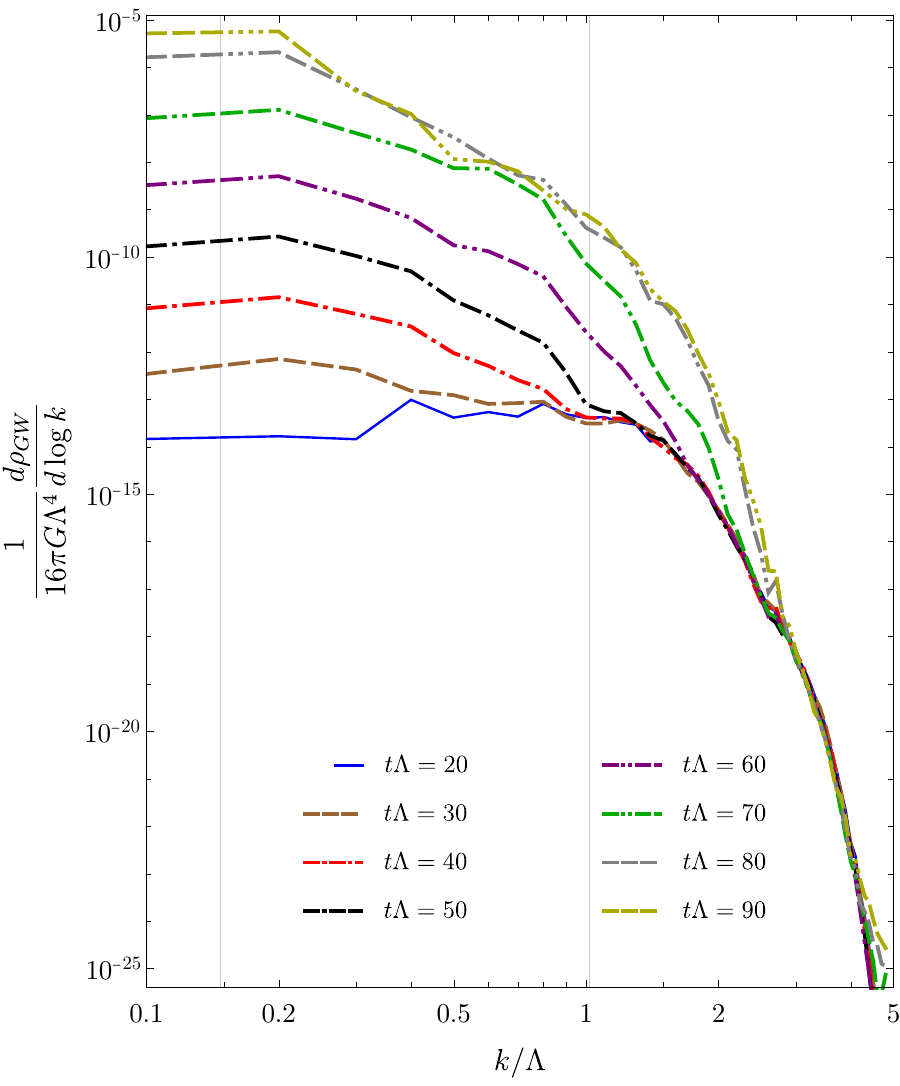}
			\caption{\label{drhoGWdlogk_linear_regime_same_a4} \small GW differential energy density per unit logarithmic momentum at early times. From left to right, the grey, vertical lines lie at $\klow \simeq 0.15\Lambda$ (computed according to \eqn{klow} with $\Lambda t=70$) and  $2\kmax$. }
	\end{center}
\end{figure}

\begin{figure}[t!!]
\begin{center}
\includegraphics[width=.99\textwidth]{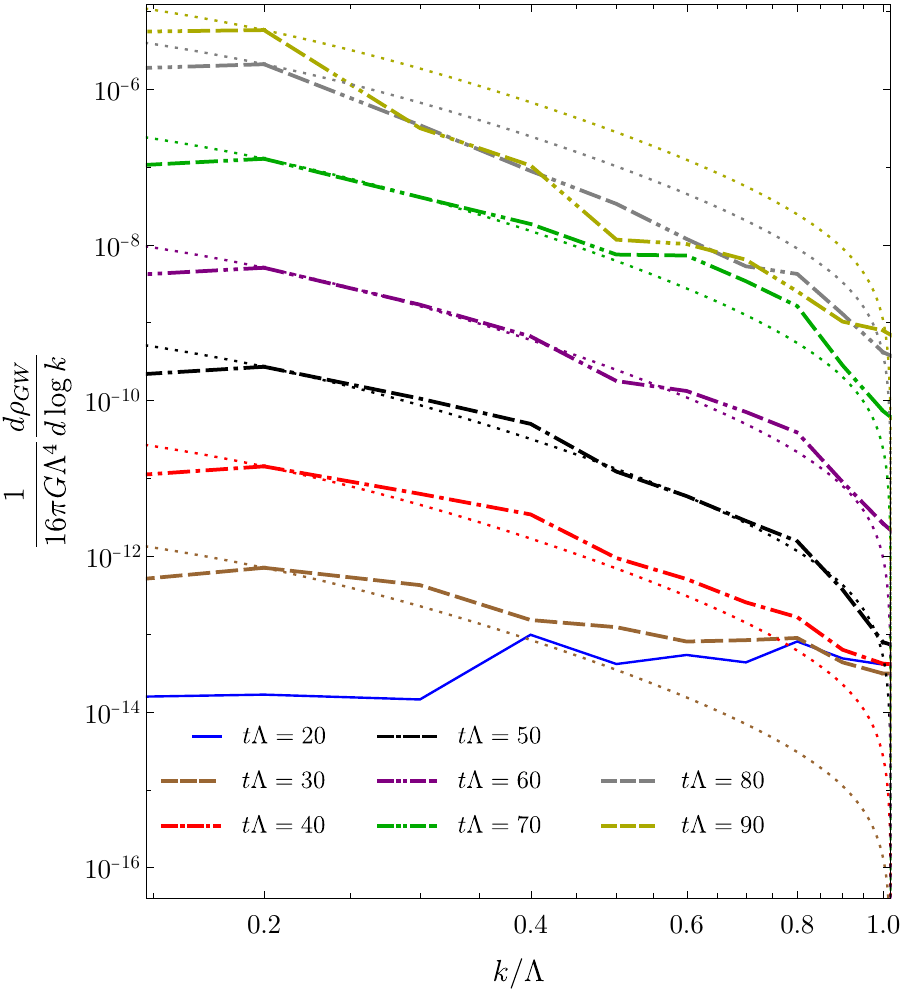}
\caption{\label{drhoGWdlogk_linear_regime_same_a4_intermediate_k} \small GW differential energy density per unit logarithmic momentum at early times for momenta approximately in the range \eqn{rangerange}. For each color, the dotted curve corresponds to the $k$-dependence predicted by \eqn{valid}, except for an overall factor that we fix by requiring agreement at $k/\Lambda=0.2$. }
	\end{center}
\end{figure}

The differential energy density per unit logarithmic momentum is shown in \figs{drhoGWdlogk_linear_regime_same_a4} and \ref{drhoGWdlogk_linear_regime_same_a4_intermediate_k}. 
In \fig{drhoGWdlogk_linear_regime_same_a4} we have marked the two values corresponding to the range \eqn{rangerange}. In \fig{drhoGWdlogk_linear_regime_same_a4_intermediate_k}  we zoom into this range. At each time we show with a dotted curve in the same color the $k$-dependence predicted by \eqn{valid}, except for an overall factor that we fix by requiring agreement at $k/\Lambda=0.2$. The need to fix this factor comes from the fact that the real evolution includes effects that were not included in the derivation of \eqn{valid}, such as initial excitations of quasi-normal modes or the decaying exponential in \eqn{decaying}. Once this factor is fixed, we see that \eqn{valid} agrees fairly well with the exact result for intermediate values of $k$ at the times around the end of the linear regime, namely for 
$40 \lesssim \Lambda t\lesssim 70$. For  values $k>2\kmax$ we see in \fig{drhoGWdlogk_linear_regime_same_a4} that the energy density decreases exponentially in $k$, as expected. We will comment on the behaviour at $k<\klow$ in \Sec{disc}.

\begin{figure}[t!!]
\begin{center}
\includegraphics[width=.99\textwidth]{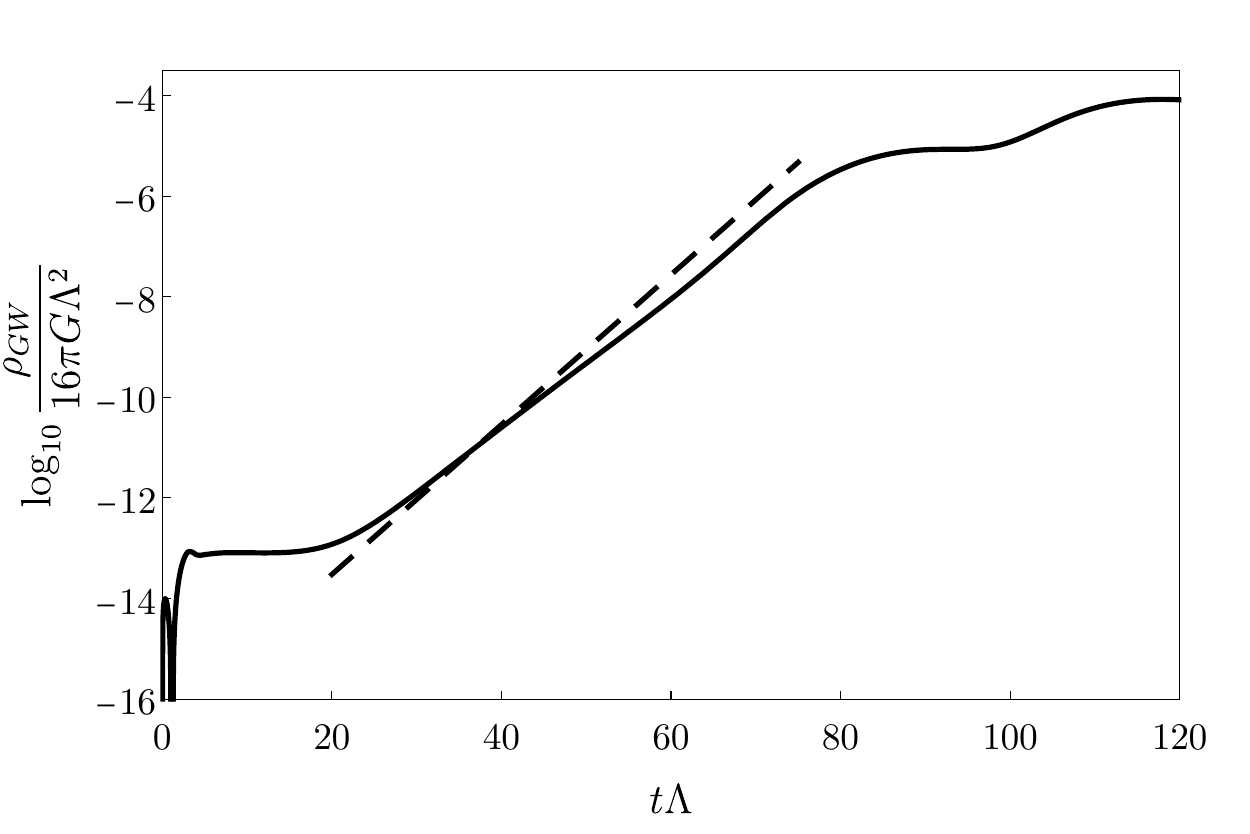}
			\caption{\label{rhoGW_log_same_a4} \small Total energy density radiated into GWs at early times. The black, dashed line corresponds to 
$\rhogw\propto e^{4\gmax t}$. 
}
	\end{center}
\end{figure}

Performing the integral over $k$ in \eqn{difTwo} or \eqn{dif} we obtain the total energy density radiated into GWs. The result at early times is shown in \fig{rhoGW_log_same_a4}. 
We see that the exponential growth of the individual modes with $k<2\kmax$ results in an analogous growth of the total energy, as indicated by the dashed line. 

\begin{figure}[t!!]
\begin{center}
\includegraphics[width=.99\textwidth]{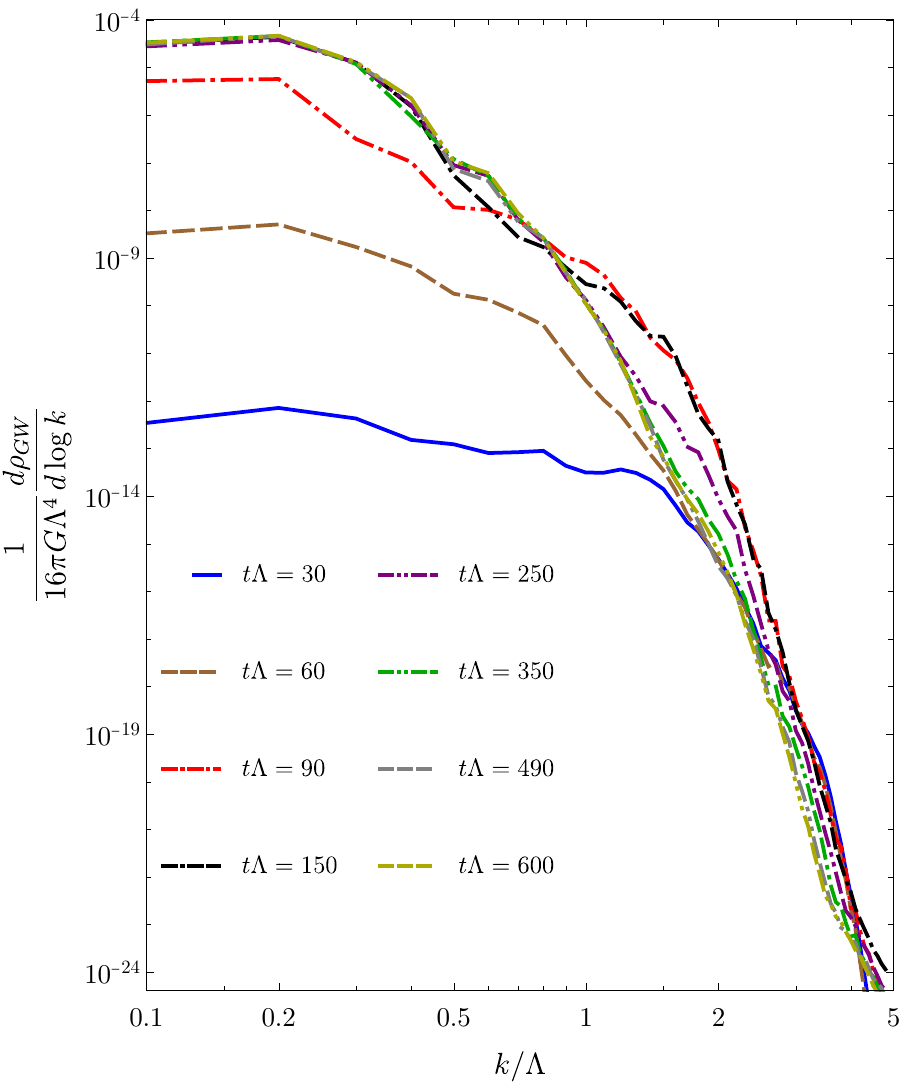}
			\caption{\label{drhoGWdlogk_same_a4} \small GW differential energy density per unit logarithmic momentum over the entire evolution. }
	\end{center}
\end{figure}

\begin{figure}[h!!]
\begin{center}
\includegraphics[width=.99\textwidth]{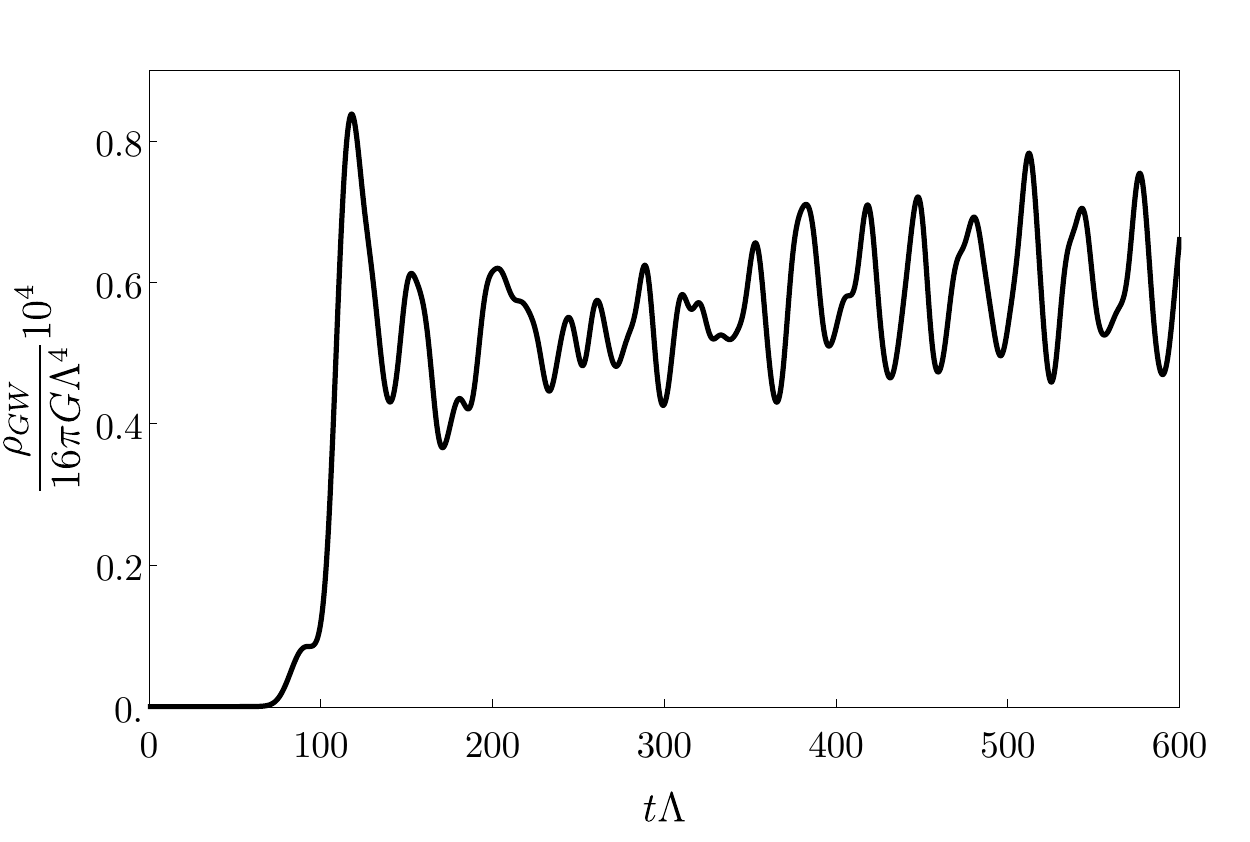}
			\caption{\label{rhoGW_same_a4} \small GW total energy density over the entire evolution. }
	\end{center}
\end{figure}

The exponential growth ceases when the non-linear regime begins around $t\Lambda \sim 70$. \fig{drhoGWdlogk_same_a4} shows the differential energy density for the entire evolution. We see that the energy density keeps increasing in the non-linear phase until it saturates at late times. 
\fig{rhoGW_same_a4} shows the corresponding total energy density on a linear scale.
In this figure we can see the oscillations in the late time behaviour expected for the energy density of a wave. We also note that most of the energy is radiated in the short reshaping period \cite{Attems:2019yqn} that takes place right after the linear regime, around $100 \lesssim \Lambda t \lesssim 150$. In this period the energy density increases from about 0.1 to about 0.7 on the scale of the plot. We will come back to this point in \Sec{disc}.

\section{Discussion}
\label{disc}
Cosmological, first-order, thermal phase transitions are usually assumed to take place via the nucleation of bubbles of the stable phase inside the metastable phase. However, if the nucleation rate is sufficiently suppressed, then the Universe can cool down all the way to the end of the metastable phase and enter the spinodal region. Under these circumstances the transition proceeds via the exponential growth of unstable modes and the subsequent formation, merging and relaxation  of phase domains. We have performed the first calculation of the GW spectrum produced by this mechanism. 

The conditions for the nucleation rate to be sufficiently suppressed depend on whether the transition takes place in the sector driving the expansion of the Universe or in a hidden sector that couples weakly, in some cases only gravitationally, to the former. The crucial point is that, as a result of the different amounts of energy injected into each sector by the reheating process, the temperature of the hidden sector may be parametrically smaller than that in the sector driving the expansion.  For concreteness we have assumed that both sectors are described by a non-Abelian gauge theory. Under these circumstances, we have shown that the constraint on the rank of the gauge group is fairly stringent if the transition takes place in the sector driving the expansion of the Universe, whereas it is fairly weak if it takes place in a hidden sector. 

Once the system enters the spinodal region, the physics at early times can be understood via a linear analysis around the unstable state. This makes two characteristic predictions. First, an exponential growth in time of the energy radiated into GWs, as illustrated by \fig{rhoGW_log_same_a4}. Second, 
as shown by \figs{drhoGWdlogk_linear_regime_same_a4} and \ref{drhoGWdlogk_linear_regime_same_a4_intermediate_k}, a specific  spectrum 
for the differential energy density per unit logarithmic momentum, $d\rhogw/d\log k$. 
Note from \eqn{tc} that, up to a factor of order unity, the scale $\Lambda$ in this and other figures coincides with the critical temperature of the phase transition, $T_c$.

Interestingly, most of the radiated energy is produced in the short reshaping period that takes place right after the linear regime, as illustrated by \fig{rhoGW_same_a4}.
Presumably, the intuition for this is as follows. When a lump of energy is accelerated, the amount of GW radiation that is emitted increases both with the amount of energy in the lump and with the acceleration. During most of the linear regime the acceleration is large because of the exponential dependence in time, but the amount of energy being displaced is small since the initial state is homogeneous. 
During this regime, peaks and valleys of increasing height and depth, respectively,  get formed. The end of the linear regime takes place precisely when these peaks and valleys are close to their maximum values. At this point, the accelerations involved in reshaping these structures are still large. This is the phase in which most of the GW emission takes place. At later times the energies being displaced are still large but the acceleration decreases as the system approaches the final, equilibrium state. 

We are currently improving the simulation presented in this paper in two ways. First, we are running simulations on bigger boxes. This is important in order to be able to explore the low-$k$ behaviour. Since the TT stress tensor attains a non-zero limit as $k\to 0$, we expect that in this limit 
\be
\frac{d\rhogw}{d\log k} \sim k^2 \,.  
\ee
At the moment our infrared cutoff is $2\pi/L\simeq 0.1\Lambda$, which is not sufficiently smaller than $\klow$ to verify this expected behaviour. Second, in our simulation we assumed that all modes have equal amplitudes at $t=0$. In order to simulate a stochastic background, we are running several simulations in which the amplitudes of these fluctuations follow a normal distribution. This Gaussianity is justified by the fact that, in a large-$N$ gauge theory, $n$-point connected correlators with $n>2$ are $1/N$-suppressed. In addition, since the second central moment of the random distribution is also $1/N$-suppressed with respect to the square of the mean energy density, the variance of the Gaussian distribution is also small in the large-$N$ limit. This motivated our choice of a small $\delta \E/\E=10^{-4}$ in the simulation presented here. Our preliminary results indicate that the stochasticity does not change our conclusions at the qualitative level. 

In our time evolution we assumed that the boundary geometry where the gauge theory lives is flat space. In other words, we ignored the expansion of the Universe. The growth rate of the unstable modes in the linear regime is comparable to the Hubble rate, see \eqq{when}, and the subsequent non-linear dynamics is slower, but not parametrically slower. Therefore, while it is reasonable to expect that neglecting the expansion of the Universe may provide a good approximation at the qualitative level, it would nevertheless be interesting to perform more sophisticated simulations including the expansion of the Universe. If the phase transition takes place in a hidden sector that is reacting to, but not affecting the, expansion of the Universe, then this would amount to fixing the boundary metric to be time-dependent but not dynamical, as in e.g.~\cite{Buchel:2017lhu,Buchel:2017pto,Buchel:2019qcq,Buchel:2019pjb,Casalderrey-Solana:2020vls,Giataganas:2021cwg,Buchel:2021ihu,Penin:2021sry}. If the transition takes place in the sector driving the expansion of the Universe, then a more rigorous calculation should include the backreaction of the degrees of freedom undergoing the transition on the spacetime metric, along the lines of \cite{Ecker:2021cvz}. In both cases we expect that the behaviour of the system at sufficiently late times will  be modified. The reason is that, in flat space, the system will tend at late times to a phase-separated state in which  two phases with energy densities $\Eh$ and $\El$ coexist. In this way the spinodal instability leads to a redistribution of the total energy in a given volume  from a homogeneous state with energy density $\E\simeq \Es$ to two approximately homogeneous regions with smaller volumes and with energy densities $\El < \Es < \Eh$. In an expanding Universe the energy density will keep decreasing everywhere, so the region with energy density $\Eh$ will cool down and eventually enter the spinodal phase again, leading to a repetition of the dynamics that we have discussed here. The volume of the region with energy density $\Eh$ decreases with each repetition. When this volume reaches a scale $\sim \Lambda^{-3}$ the spinodal instability no longer leads to a phase-separated configuration \cite{Attems:2019yqn} and the process stops,  leading to the completion of the phase transition. We plan to report  on the details of this process elsewhere.

\acknowledgments
It is a pleasure to thank Bartomeu Fiol, Oscar Henriksson, Mark Hindmarsh, Carlos Hoyos and Oriol Pujol\`as for discussions.  YB acknowledges support from the European Research Council Grant No. ERC-2014-StG 639022-NewNGR and the Academy of Finland grant no. 333609. TG acknowledges financial support from FCT/Portugal Grant No.\ PD/BD/135425/2017 in the framework of the Doctoral Programme IDPASC-Portugal. 
AJ acknowledges support from the European Research Council Grant No. ERC-2016-AvG 692951-GravBHs. MSG acknowledges financial support from the APIF program, fellowship APIF\_18\_19/226. JCS, DM and MSG are also supported by grants SGR-2017-754, PID2019-105614GB-C21, PID2019-105614GB-C22 and the ``Unit of Excellence MdM 2020-2023'' award to the Institute of Cosmos Sciences (CEX2019-000918-M). MZ acknowledges financial support provided by FCT/Portugal through the IF programme grant IF/00729/2015,
and CERN/FIS-PAR/0023/2019.
The authors thankfully acknowledge the computer resources, technical expertise and assistance provided by CENTRA/IST. Computations were performed in part at the cluster ``Baltasar-Sete-S\'ois'' and supported by the H2020 ERC Consolidator Grant ``Matter and strong field gravity: New frontiers in Einstein's theory'' grant agreement No.\ MaGRaTh-646597. We also thank the MareNostrum supercomputer at the BSC (activity Id FI-2021-1-0008) for significant computational resources.





\bibliography{Spinodal}{}
\bibliographystyle{JHEP}
\end{document}